\documentclass{article}

\usepackage[utf8]{inputenc}
\usepackage{multirow}
\usepackage{multicol}
\usepackage[pdftex]{graphicx}   
\usepackage{array}
\usepackage{mathtools, nccmath}
\usepackage{hyperref}
\usepackage{authblk}
\usepackage{color,soul}
\usepackage{rotating}
\definecolor{darkgreen}{rgb}{0.0, 0.5, 0.13}
\definecolor{orange}{rgb}{1, 0.5, 0}
\newcommand*\rot{\rotatebox{90}}

\newcolumntype{R}[1]{>{\raggedleft\let\newline\\\arraybackslash\hspace{0pt}}m{#1}}

\begin{document}

\title{Real-time malware process detection and automated process killing}

\author[1,2]{Matilda Rhode\thanks{This research was part funded by the Engineering and Physical Sciences Research Council (EPSRC) – grant references EP/P510452/1 and EP/S035362/1. The research was also part funded by Airbus Operations Ltd. Data is available from the corresponding author on request}}
\author[2]{Pete Burnap}
\author[1]{Adam Wedgbury}
\affil[1]{\{matilda.rhode, adam.wedgbury\}@airbus.com \authorcr Airbus, Newport, UK}
\affil[2]{{BurnapP}@cardiff.ac.uk \authorcr Cardiff University, UK}

\date{}                     %% if you don't need date to appear
\setcounter{Maxaffil}{0}
\renewcommand\Affilfont{\itshape\small}

\maketitle

\begin{abstract}

Perimeter-based detection is no longer sufficient for mitigating the threat posed by malicious software. This is evident as antivirus (AV) products are replaced by endpoint detection and response (EDR) products, the latter allowing visibility into live machine activity rather than relying on the AV to filter out malicious artefacts. This paper argues that detecting malware in real-time on an endpoint necessitates an automated response due to the rapid and destructive nature of some malware. 

The proposed model uses statistical filtering on top of a machine learning dynamic behavioural malware detection model in order to detect individual malicious processes on the fly and kill those which are deemed malicious. In an experiment to measure the tangible impact of this system, we find that fast-acting ransomware is prevented from corrupting 92\% of files with a false positive rate of 14\%. Whilst the false-positive rate currently remains too high to adopt this approach as-is, these initial results demonstrate the need for a detection model which is able to act within seconds of the malware execution beginning; a timescale that has not been addressed by previous work. 

\end{abstract}

\section{Introduction}\label{sec:introduction}

Our increasingly digitised world broadens both the opportunities and motivations for cyber attacks, which can have devastating social and financial consequences \cite{tatar2021digital}. Malicious software (malware) is one of the most commonly used vectors to propagate malicious activity and exploit code vulnerabilities. 

Due to the huge numbers of new malware appearing each day, the detection of malware samples needs to be automated \cite{sahay2020evolution}. Signature-matching methods are not resilient enough to handle obfuscation techniques nor to catch unseen malware types and as such, automated methods of generating detection rules, such as machine learning have been widely proposed e.g. \cite{Huang2016MtNet, hu2018black, chen2019gene, Ijaz2019StaticAndDynamic}. These approaches typically analyse samples when the file is first ingested, either using static code-based methods or by observing dynamic behaviours in a virtual environment. 

This paper argues that both of these approaches are vulnerable to evasion from the attacker. Static methods may be thwarted by simple code-obfuscation techniques whether rules are hand-generated \cite{obfuscationTechniques} or created using machine learning \cite{kolosnjaji2018adversarial}. Dynamic detection in a sandboxed environment cannot continue forever, either it is time-limited e.g. \cite{carlin2019cost} or ends after some period of inactivity e.g. \cite{ShibaharaEfficientDynamics}. This fixed period allows attackers to inject benign activity during analysis and wait to carry out malicious activity once the sample has been deemed harmless and passed on to the victim's environment. 
The pre-execution filtering of malware is the model used by antivirus but this is insufficient to keep up with the ever-evolving malware landscape and has lead to the creation of endpoint detection and response (EDR) products which allow security professionals to monitor and respond to malicious activity on the victim machine. Real-time malware detection also monitors malware live on the machine thus capturing any malicious activity on the victim machine even if it was not evident during initial analysis. This paper proposes that once a threat is detected, due to the fast-acting nature of some destructive malware, it is vital to have automated actions to support these detections. In this paper we investigate automated detection and killing of malicious processes for endpoint protection. 

There are several key challenges to address in detecting malware on-the-fly on a machine in use by comparison with detecting malicious applications that are detonated in isolation in a virtual machine. These are summarised below: 

\begin{enumerate}
    \item \textbf{Signal Separation:} Detection in real time requires that the malicious and benign activity are separated in order that automated actions can be taken on only the malicious processes.
    \item \textbf{Use of Partial Traces:} In order to try and mitigate damage, malware needs to be detected as early as possible but, as shown in previous work \cite{rhode2018early}, there is a trade-off between the amount of data collected and classification accuracy in the first few seconds of an application launching and the same may be true for individual processes.
    \item \textbf{Quick Classification:} The inference itself should be as fast as possible in order to further limit the change of malicious damage once the process is deemed malicious.
    \item \textbf{Impact of Automated Killing in Supervised Learning:} Supervised learning averages the error rate across the entire training set but when the classification results in an action, this smoothing out of errors across the temporal dataset is not possible.
\end{enumerate}

This paper seeks to address these key challenges and provides preliminary results including a measure of 'damage prevented' in a live environment for fast-acting destructiveware. As well as the results from these experiments this paper contributes an analysis of the computational resources against detection accuracy for many of the most popular machine learning algorithms used for malware detection.

The key contributions of this paper are as follows:
\begin{itemize}
    \item The first general malware detection model to demonstrate damage mitigation in real-time using process detection and killing
    \item Benchmarking of commonly used ML algorithm implementations with respect to computational resource consumption
    \item Presentation of real-time malware detection against more user background applications than have previously been investigated; increasing from 5 to 36 (up to 95 simultaneous processes)
\end{itemize}

The next section outlines related work, followed by a report of the three methodologies that were tested to try and address these challenges \ref{methodology} in which the method for evaluating these models is also explained (\ref{sec:4_ransomware}). The experimental setup is described in section~\ref{sec:4_environment} followed by results in sections~\ref{sec:4_results}. 

\section{Related Work}

\subsection{Malware detection with static or post-collection behavioural traces}

\textbf{Static sources:} Machine learning models trained on static data have shown good detection accuracy, e.g. Chen et. al.\cite{chen2019gene} achieved 96\% detection accuracy using statically-extracted sequences of API calls to train a Random Forest model. However, static data has been demonstrated to be quite vulnerable to concept drift \cite{saxe2015deep, pendlebury2019tesseract}. Adversarial samples present an additional emerging concern; Grosse et al.\cite{grosse2016adversarial} and Kolosnaji et al.\cite{kolosnjaji2018adversarial} demonstrated that static malware detection models achieving over 90\% detection accuracy could be thwarted by injecting code or simply altering the padded code at the end of a compiled binary respectively. 

\textbf{Post-collection dynamic data:} Dynamic behavioural data is generated by the malware carrying out its functionality. Again machine learning models have been used to draw out patterns between malicious and benign software using dynamic data. Various dynamic data can be collected to describe malware behaviour. The most commonly used data are API calls made to the operating system, typically recorded in short sequences or by frequency of occurrence. Huang and Stokes's research \cite{Huang2016MtNet} reports the highest accuracy in recent malware detection literature with a very large dataset of more than 6 million samples to achieve an accurate detection rate of 99.64\% using a neural network trained on the input parameters passed to API calls, their return values, and the co-occurrence of API calls. Other dynamic data sources include dynamic opcode sequences (e.g. Carlin et al.\cite{carlin2019cost} achieve 99\% using a Random Forest), hardware performance counters (e.g. Sayadi \cite{sayadi2018ensemble} achieve 94\% on Linux/Ubuntu malware using a decision tree), network activity and file system activity (e.g. Usman et al. \cite{usman2021intelligent} achieve 93\% using a decision tree in combination with threat intelligence feeds and these data sources), and machine activity metrics (e.g. Burnap et al. \cite{somburnap} achieve 94\% using a self-organising map). Previous work \cite{rhode2019lab} demonstrated the robustness of machine activity metrics over API calls in detecting malware collected from different sources. 

Dynamic detection is more difficult to obfuscate but typically the time taken to collect data is several minutes, making it less attractive for endpoint detection systems. Some progress has been made on early detection of malware. Previous work \cite{rhode2018early}) was able to detect malware with 94\% accuracy within 5 seconds of execution beginning. However, as a sandbox-based method, malware which is inactive for the first 5 seconds is unlikely to be detected with this approach. Moreover, the majority of dynamic malware detection papers use virtualised environments to collect data.

\subsection{Real-time malware detection with partial behavioural traces}

\begin{table}[!ht]
    \centering
    \small
    \begin{tabular}{c|p{0.3cm}|p{0.3cm}|p{0.3cm}|p{0.3cm}|p{0.3cm}|p{0.3cm}|p{1.3cm}|p{1.3cm}|R{0.8cm}|p{1.5cm}|p{1.5cm}}
        & \multicolumn{5}{c|}{Problem considered} \\\hline
       Ref.    & \rot{(1) Signal separation} & \rot{(2) Early detection} & \rot{(3) Quick Classification / Latency} & \rot{(4) Impact of automated actions} & \rot{Resource consumption} & \rot{Real-time tested} & \rot{Malware types} & \rot{OS} & \rot{\# Samples} & \rot{Features} & \rot{Algorithm} \\\hline
        \cite{sayadi2018comprehensive}  & & & X & & X & & General & Linux & 200 & HPCs & Boosted DT\\\hline
        \cite{sayadi2018ensemble} & & & X & & X & & General & Linux & 200 & HPCs & Boosted DT \\\hline
        \cite{das2016semantics}  & & X & X & & X & X & General & Linux & 798 & API calls & MLP \\\hline
        \cite{ozsoy2015hardware} & & X & & & X & X & General & Windows & 1,554 & memory addresses, instructions &  NN  \\\hline
        \cite{yuan2017phd}  & & & X & & & X & General & Windows, Linux & 500 & API calls & NN \\\hline
        \cite{fastSlow2017}  & X & X & X & & & X & General & Windows & 9,992 & API calls & RF + NN \\\hline
        \cite{scaife2016cryptolock}  & & X & X & X & & X & Crypto Ransomware & Windows & 497 & File data & Rules \\\hline
    \end{tabular}
    \caption{Real-time malware detection literature problems considered. OS = operating system; HPCs = Hardware performance counters; DT=Decision Tree; MLP = multi-layer perceptron; NN = Neural Network; RF=Random Forest}
    \label{tab:1_real-time_lit}
\end{table}

Previous work has begun to address the four challenges set out in the introduction. Table ~\ref{tab:1_real-time_lit} summarises the related literature and the problems considered by the researchers.

To the best of our knowledge, challenge \textit{(1) signal separation} has only previously been addressed by Sun et al. \cite{fastSlow2017} using sequential API call data. The authors execute up to 5 benign and malicious programs simultaneously achieving 87\% detection accuracy after 5 minutes of execution and 91\% accuracy after 10 minutes of execution. 

Challenge \textit{(2) to detect malware using partial traces as early as possible} has not been directly addressed. Some work has looked at early run-time detection; Das et al. \cite{das2016semantics} used an FPGA as part of a hybrid hardware-software approach to detect malicious Linux applications using system API calls which are then classified using a multilayer perceptron. Their model was able to detect 46\% of malware within the first 30\% of its execution with a false-positive rate of 2\% in offline testing. These findings however were not tested with multiple benign and malicious programs running simultaneously and do not explain the impact of detecting 46\% of malware within 30\% of its execution trace in terms of benefits to a user or the endpoint being protected. How long does it take for 30\% of the malware to execute? What has occurred in that time?

Greater attention has been paid to challenge \textit{(3) quick classification}, insofar as this problem also encompasses the need for lightweight detection. Some previous work has proposed hardware based detection for lightweight monitoring. Syadi et al. \cite{sayadi2018ensemble} use high performance counters (HPCs) as features to train ensemble learning algorithms and scored 0.94 AUC using a dataset of 100 malicious and 100 benign Linux software samples. Ozsoy et al. \cite{ozsoy2015hardware} use low-level architectural events to train a multilayer perceptron on the more widely used \cite{win7} (and attacked) Windows operating system. The model was able to detect 94\% of malware with a false positive rate of 7\% using partial execution traces of 10,000 committed instructions. The hardware based detection models however, are less portable than software-based systems due to the ability for the same operating system to run on a variety of hardware configurations. 

Both Sun et al. \cite{fastSlow2017} and Yuan \cite{yuan2017phd} propose two-stage models to address the need for lightweight computation. The first stage comprises a lightweight ML model such as a Random Forest to alert suspicious processes, the second being a deep learning model which is more accurate but more computationally intensive to run. Two-stage models, as Sun et al. \cite{fastSlow2017} note, can get stuck in an infinite loop of analysis in which the first model flags a process as suspicious but the second model deems it benign and this labelling cycle continues repeatedly. Furthermore, if the first model of the two is prone to false negatives, malware will never be passed to the second model for deeper analysis.

Challenge \textit{(4) the impact of automated actions} has been discussed by Sun et al. \cite{fastSlow2017}. The authors also propose the two-stage approach as a solution to this problem. The authors apply restrictions to the process whilst the deeper NN analysis takes place followed by the killing of malicious-labelled processes. The authors found that the delaying strategy impacted benignware more than malware and used this two-stage process to account for the irreversibility of the decision to kill a process. The authors did not assess the impact on the endpoint with respect to the time at which the correctly classified malware was terminated.

\section{Methodology - three approaches}\label{methodology}

As noted above, supervised learning models average errors across the training set but in the case of real-time detection and process killing, a \textit{single} false positive on a benign process amongst 300 true-negatives would cause disruption to the user. The time at which an malware is detected is also important, the earlier the better. Therefore the supervised learning model needs to be adapted to take account of these new requirements. 

Tackling this issue was attempted in three different ways and all three are reported here in the interests of reporting negative results as well as the one which performed the best. These were:

\begin{enumerate}
    \item Statistical methods to smooth the alert surface and filter out single false-positives
    \item Reinforcement learning, which is capable of incorporating the consequences of model actions into learning
    \item A regression model based on the feedback of a reinforcement learning model made possible by having the ground-truth labels
\end{enumerate}

\begin{figure}[!htpb]
    \centering
    \includegraphics[width=\textwidth]{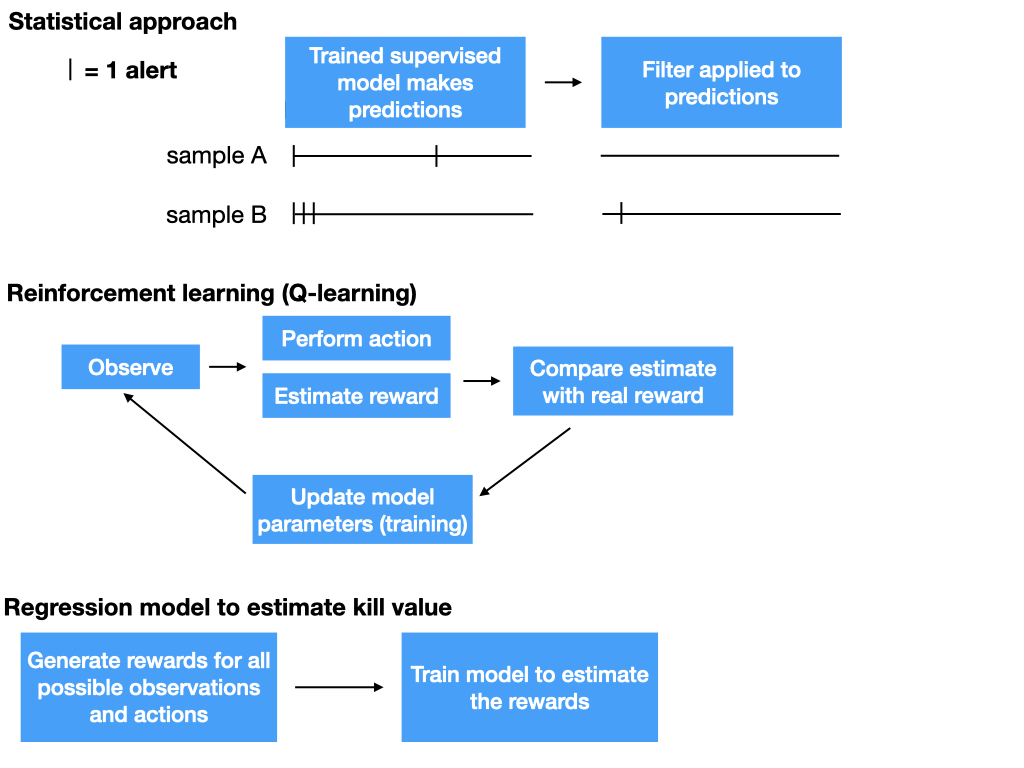}
    \caption{High-level depiction of three approaches taken}
    \label{fig:4_approaches}
\end{figure}

Figure~\ref{fig:4_approaches} gives a high-level depiction of the three approaches tested in this paper. 
 
\subsection{Statistical Approach: Alert Filtering}

It is expected that transitioning from a supervised learning model to a real-time model will see a rise in false-positives since one single alert means benign processes (and all child processes) are terminated, which effectively renders all future data points as false positives. Filtering the output of the models, just as the human brain filters out transient electrical impulses in order to separate background noise from relevant data \cite{benda2005spike}, may be sufficient to make supervised models into suitable agents. This is attractive because supervised learning models are already known to perform well for malware detection, as confirmed by the previous paper and other related work \cite{rhode2018early, huang2011malware, das2016semantics, kim2018runtime}. A disadvantage of this approach is that it introduces additional memory and computational requirements both in order to calculate the filtered results and to track processes current and historic scores, therefore a model which integrates the expected consequences of an action into learning is also tested: reinforcement learning.

\subsection{Reinforcement Learning: Q-learning with Deep Q Networks}

The proposed automated killing model may be better suited to a reinforcement learning strategy than to supervised learning. Reinforcement learning uses rewards and penalties from the model's environment. The problem that this paper is seeking to solve is essentially a supervised learning problem, but one for which it is not possible to average predictions. There are no opportunities to classify the latter stages of a process if the agent kills the process, and this can be reflected by the reward mechanism of the reinforcement learning model (see Figure~\ref{fig:4_approaches} above). Therefore reinforcement learning seems like a good candidate for this problem space. 

Two limitations of this approach are that \textit{(1)} reinforcement learning models can struggle to converge on a balanced solution, the models must learn to balance the exploration of new actions with the re-use of known high-reward actions; commonly known as the exploration-exploitation trade-off \cite{sutton1998introduction} \textit{(2)} in these experiments, the reward is based on the malware/benignware label at the application level rather than being linked to the actual damage being caused, therefore the signal is a proxy for what the model should be learning. This is used because, as discussed above, the damage caused by different malware is subjective.

For reinforcement learning, loss functions are replaced by reward functions which update the neural network weights to reinforce actions (in context) that lead to higher rewards and discourage actions (in context) that lead to lower rewards; these contexts and actions are known as state-action pairs. Typically the reward is calculated from the perceived value of the new state that the action leads to e.g. points scored in a game. Often this cannot be pre-labelled by a researcher since there are so many (maybe infinite) state-action pairs. However in this case, all possible state-action pairs can be enumerated, which is the third approach tested (regression model - outlined in the next section).

The reinforcement model was still tested. Here the reward is \(+N\) for a correct prediction, \(-N\) for an incorrect prediction where \(N\) is the total number of processes impacted by the prediction e.g. if there is only one process in a process tree but 5 more will appear over the course of execution, a correct prediction gives a reward of \(+6\), and incorrect prediction gives a reward of \(-6\). 

There are a number of reinforcement learning algorithms to choose from. This paper explores q-learning \cite{watkins1989learning, watkins1992q, sutton1990integrated, lin1992self} to approximate the value or 'quality' (q) of a given action in a given situation. Q-learning approximates q-tables, which are look-up tables of every state-action pair and their associated rewards. A state-action pair is a particular state in the environment coupled with a particular action i.e. the machine metrics of the process at a given point in time with the action to leave the process running. When the number of state-action pairs becomes quite large, it is easier to approximate the value using an algorithm. Deep Q networks (DQN) are neural networks that implement q-learning and have been used in state-of-the-art reinforcement learning arcade game playing, see Mnih et al. \cite{mnih2013playing}. A DQN was the reinforcement algorithm trialled here, though it did not perform well by comparison with the other methods, a different RL algorithm may perform better \cite{mnih2016asynchronous}, but the results are still included in the interests of future work. The following paragraphs will explain some of the key features of the DQN.

The DQN tries out some actions, stores the states, actions, resulting states and rewards in a memory and uses these to learn the expected rewards of each available action; with the highest expected reward being the one that is chosen. Neural networks are well-suited to this problem since their parameters can easily be updated, tree-based algorithms like random forests and decision trees can be adapted to this end but not as easily. Future rewards can be built into the reward function and are usually discounted according to a tuned parameter usually signified by \(\gamma\).

In Mnih et al's \cite{mnih2013playing} formulation, in order to address the exploration-exploitation trade off, DQNs either exploit a known action or explore a new one, with the chance of choosing exploration falling over time. When retraining the model based on new experiences, there is a risk that previous useful learnt behaviours are lost, this problem is known as catastrophic forgetting \cite{kirkpatrick2017overcoming}. Mnih et al's \cite{mnih2013playing} DQNs use two tools to combat this problem. First, experience replay by which past state-action pairs are shuffled before being used for retraining so that the model does not catastrophically forget. Second, DQNs utilise a second network, which updates at infrequent intervals in order to stabilise the learning.

Q-learning may enable a model to learn when it is confident enough to kill a process, using the discounted future rewards. For example, choosing not to kill some malware at time \(t\) may have some benefit as it allows the model to see more behaviour at t+1 which gives the model greater confidence that the process is in fact malicious. Whilst there are other reinforcement learning algorithms. 

Q-learning approximates rewards from experience, but in this case, all rewards from state-action pairs can actually be pre-calculated. Since one of the actions will kill the process and thus end the `experience' of the DQN, it could be difficult for this model to gain enough experience. Thus pre-calculation of rewards may improve the breadth of experience of the model, for this reason a regression model is proposed to predict the Q-value of a given action. 

\subsection{Regression using Q-Values}

Unlike classification problems, regression problems can predict a continuous value rather than discrete (or probabilitic) values relating to a set of output classes. Regression algorithms are proposed here to predict the q-value of killing a process. If this value is positive, the process is killed. 

Q-values estimate the value of a particular action based on the `experience' of the agent. Since the optimal action for the agent is always known, it is possible to precompute the `(q-)value' of killing a process and train various ML models to learn this value. It would typically be quicker to train a regression model which tries to learn the value of killing a process than to train a DQN which explores the state-action space and calculates rewards between learning, since the interaction and calculation of rewards is no longer necessary. The regression approach can be used with any machine learning algorithm capable of learning a regression problem, regardless of whether it is capable of partial training. 

There are two primary differences between this regression approach and the reinforcement learning DQN approach detailed in the previous section. Firstly, the datasets are likely to be difference. Since the DQN generates training data through interacting with its environment it may never see certain parts of the state-action space e.g. if a particular process \(A\) is always killed during training before time \(t*\), the model is not able to learn from the process \(A\) data after \(t*\).

Secondly, \textit{only} the expected value of killing is modelled by the regressor, whereas the DQN tries to predict the value of both killing and of not killing the process. This means that the equation used to model the value of process killing is only an approximation of the reward function used by the DQN.

The equation used to calculate the value of killing is positive for malware and negative for benignware, in both cases it is scaled by the number of child processes impacted and in the case of malware, early detection increases the value of process killing (with an exponential decay). Let \(y\) be the true label of the process (0=benign, 1=malicious), \(N\) is the number of child processes and \(t\) is the time in seconds at which the process is killed then the value of killing a process is:

\[(y * 2 - 1) * (1 + N) * (1 + (y * (e^{-t})))\]

The equation above negatively scores the killing of benignware in proportion to the number of subprocesses and scores the killing of malware positively in proportion to the number of subprocesses. A bonus reward is scored for killing malware early, with an exponential decay over time.

\section{Evaluation Methodology: Ransomware detection}\label{sec:4_eval}

As noted in the background paper, to date research has not addressed the extent to which damage is mitigated by process killing, since Sun et al. \cite{fastSlow2017} presented the only previous work to test process killing and damage with and without process killing is not assessed. To this end, this paper uses ransomware as a proxy to detect malicious damage, inspired by Scaife et. al's approach \cite{scaife2016cryptolock}. A brief overview of Scaife et al.'s damage measurement is outlined below:

Early detection is particularly useful for types of malware from which recovery is difficult and/or costly. Cryptographic ransomware encrypts user files and withholds the decryption key until a ransom is paid to the attackers. This type of attack is typically costly to remedy, even if the victim is able to carry out data recovery \cite{soposRansomware}. Scaife et al.'s work \cite{scaife2016cryptolock} on ransomware detection uses features from file system data, such as whether the contents appears to have been encrypted, and number of changes made to the file type. The authors were able to detect and block all of the 492 ransomware samples tested with less than 33\% of user data being lost in each instance. Continella et al. \cite{continella2016shieldfs} propose a self-healing system, which detects malware using file system machine activity (such as read/write file counts), the authors were able to detect all 305 ransomware samples tested, with a very low false-positive rate. These two approaches use features selected specifically for their ability to detect ransomware, but this requires knowledge of how the malware operates. Whereas the approach taken here seeks to use features which can be used to detect malware \textit{in general}. The key purpose of this final experiment (section ~\ref{sec:4_ransomware}) is to show that our general model of malware detection is able to detect general types of malware as well as time-critical samples such as ransomware.

\section{Experimental Setup}\label{setup}

This section outlines the data capture process and dataset statistics.

\subsection{Features}

The same features as were used in previous work \cite{rhode2018early} are used here for process detection, with some additional features to measure process-specific data. Despite the popularity of API calls noted in \cite{rhode2019lab}, due to these findings and Sun et al.'s \cite{fastSlow2017} difficulties hooking this data in real-time, these were not considered as features to train the model. 

\begin{table}[ht!]
\small
\centering
\begin{tabular}{
R{0.2\textwidth}|
R{0.2\textwidth}|
R{0.2\textwidth}|
R{0.2\textwidth}}
\textbf{Category} \\\hline\hline
\textbf{CPU use} (\%) & system level & user level \\\hline
\textbf{Memory use} (bytes) & total & physical (non-swapped) & swap \\\hline
\textbf{Child processes} & count & maximum process ID & number of threads\\\hline
\textbf{I/O operation bytes on disk} (bytes) & read & write & non-read-write I/O operations \\\hline
\textbf{I/O operation count on disk} & read & write & non-read-write I/O operations \\\hline
\textbf{Priority} & process priority & I/O process priority\\\hline
\textbf{Network \# Packets} & TCP packet count & UDP packet count  \\\hline
\textbf{Network \# Bytes} & \# bytes sent & \# bytes received  \\\hline
\textbf{Network Other} & number of connections currently open & statuses of the ports opened by the process (4 statuses) \\\hline
\textbf{Miscellaneous} & number of command line arguments passed to process &  number of handles being used by process\\\hline
\end{tabular}
\caption{26 process-level features: 22 features + 4 port status values}
\label{tab:c3_features}
\end{table}

At the process-level, 26 machine metric features are collected, these were dictated by the attributes available using the Psutil \cite{psutil} python library. It is also possible to include the `global' machine learning metrics that were used in the previous papers. Though global metrics will not provide process-level granularity, they may give muffled indications of the activity of a wider process tree. The 9 global metrics are: system level CPU use, user level CPU use, memory use, swap memory use, number of packets received and sent, number of bytes received and sent and the total number of processes running.

The process-level machine activity metrics collected are: CPU use at the user level, CPU use at the system level, physical memory use, swap memory use, total memory use, number of child process, number of threads, maximum process ID from a child process, disk read, write and other I/O count, bytes read, written and used in other I/O processes,  process priority, I/O process priority, number of command line arguments passed to process, number of handles being used by process, time since the process began, TCP packet count, UDP packet count, number of connections currently open, 4 port statuses of those opened by the process (see Table~\ref{tab:c3_features}). 

\subsubsection{Preprocessing} Feature normalisation is necessary for NNs to avoid over-weighting features with higher absolute values. The test, train and validation sets (\(x\)) are all normalised by subtracting the mean (\(\mu\)) and dividing by the standard deviation (\(\sigma\)) of each feature in the training set: \(\frac{x - \mu}{\sigma}\). This sets the range of input values largely between -1 and 1 for all input features, avoiding the potential for some features to be weighted more important than others during training purely due to the scalar values of those features. This requires additional computational resources but is not necessary for all ML algorithms; this is another reason why the supervised RNN used in \cite{rhode2018early} may not be well-suited for real-time detection.

\subsection{Data Capture}

During data capture, this research sought to improve upon previous work and emulate real machine use to a greater extent than has previously been trialled. The implementation details of the VM,  simultaneous process execution and RL simulation are outlined below:

\subsubsection{Environment: Machine setup}\label{sec:4_environment}
The following experiments were conducted using a virtual machine (VM) running with Cuckoo Sandbox \cite{cuckooSandbox} for ease of collecting data and restarting between experiments and because the Cuckoo Sandbox emulates human interaction with programs to some extent to promote software activity. In order to emulate the capabilities of a typical machine, the modal hardware attributes of the top 10 `best seller' laptops according to a popular internet vendor \cite{amazon} were used, and these attributes were the basis of the VM configuration. This resulted in a VM with 4GB RAM, 128GB storage and dual-core processing running Windows 7 64-bit. Windows 7 was the most prevalent computer operating system (OS) globally at the time of designing the experiment \cite{win7}, though Windows 10 is now the most popular OS, the findings in this research should still be relevant.

\subsubsection{Simultaneous applications}
In typical machine use, multiple applications run simultaneously. This is not reflected by behavioural malware analysis research in which samples are injected individually to a virtual machine for observation. The environment used for the following experiments launches multiple applications on the same machine at slightly staggered intervals as if a user were opening them. Each malware is launched with a small number (1-3) and a larger number (3-35) of applications. It was not possible to find find up-to-date user data on the number of simultaneous applications running on a typical desktop, so here it was elected to launch up to 36 applications (35 benign + 1 malicious) at once, which is the largest number of simultaneous apps for real-time data collection to date. From the existing real-time analysis literature only Sun et al. \cite{fastSlow2017} run multiple applications at the same time, with a maximum of 5 running simultaneously. 

Each application may in turn launch multiple processes, causing more than 35 processes to run at once; 95 is the largest number of simultaneous processes recorded, this excludes background OS processes. 

\subsubsection{Reinforcement Learning Simulation}
For reinforcement learning, the DQN requires an observation of the resulting state following an action. To train the model, a simulated environment is created from the pre-collected training data whereby the impact of killing or not killing a process is returned as the next state. For process-level elements this reduces all features to zero. A caveat here is that in reality killing the process may not occur immediately and therefore memory, processing power etc. may still be being consumed at the next data observation. For global metrics, the process-level values for the killed processes (includes child processes of the killed process) are subtracted from the global metrics. There is a risk again that this calculation may not correlate perfectly with what would be observed in a live machine environment. 

\begin{figure}
    \centering
    \begin{minipage}{0.9\textwidth}
        \centering
        \includegraphics[width=0.9\textwidth]{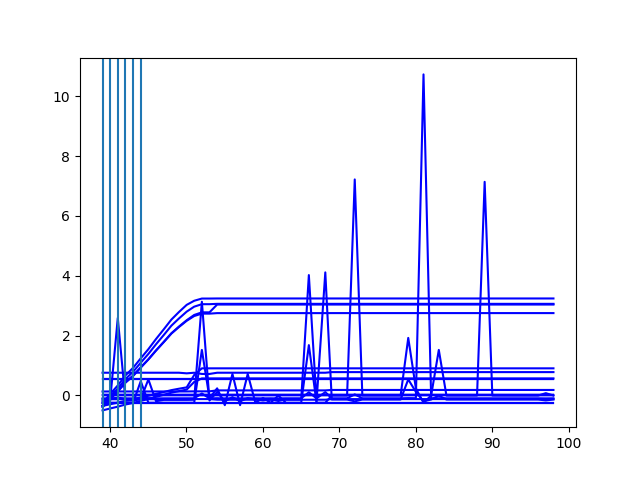} % first figure itself
        \caption{Benignware sample, normalised process-level metrics, 6 observations made without process being killed}\label{fig:B_RLenv}
    \end{minipage}\hfill
    \begin{minipage}{0.9\textwidth}
        \centering
        \includegraphics[width=0.9\textwidth]{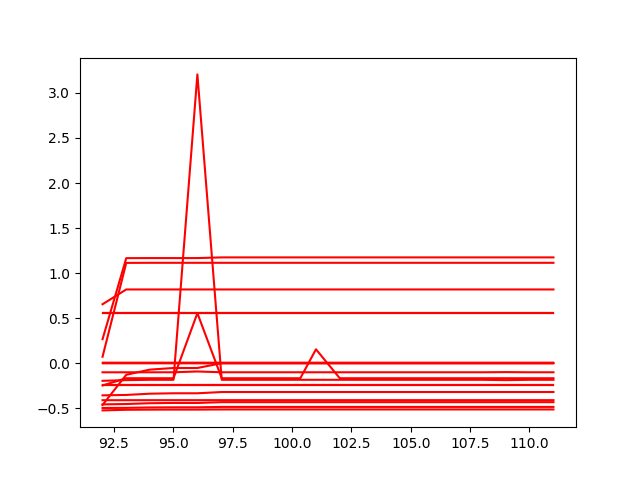}
        \caption{Malware sample, normalised process-level metrics, no observations made yet}
        \label{fig:4_RLenv}
    \end{minipage}
\end{figure}
\clearpage

In order to observe the model performance, a visualisation was developed to accompany the simulated environment. Figures~\ref{fig:B_RLenv} and \ref{fig:4_RLenv} show a screenshots of the environment visualisation for one malicious and one benign process. 

\subsection{Dataset}

The dataset is comprised of 3,604 benign executables and 2,792 malicious applications (each containing at least one executable), with 2,877 for training and validation and 3,519 for testing. These dataset sizes are consistent with previous real-time detection dataset sizes e.g. (Das et al. \cite{das2016semantics} use 168 malicious, 370 benign; Sayadi et al. \cite{sayadi2018ensemble} use over 100 each benign and malicious;  Ozsoy et al. \cite{ozsoy2015hardware} use 1,087 malicious and 467 benign; Sun et al. \cite{fastSlow2017} use 9,115 malicious, 877 benign). 
With multiple samples running concurrently to simulate real endpoint use, there are 24K processes in the training set and 34K in the test set. Overall there are 58K behavioural traces of processes in the training and testing datasets. The benign samples comprise files from VirusTotal \cite{virusTotal}, from free software websites (later verified as benign with VirusTotal), and from a fresh Microsoft Windows 7 installation. The malicious samples were collected from two different VirusShare \cite{VirusShare} repositories. 

In Pendelbury et al's analysis \cite{pendlebury2019tesseract}, the authors estimate that in the wild between 6\% and 22\% of applications are malicious, normalising to 10\% for their experiments. Using this estimation of Android malware, ratios a similar ratio was used in the test set in which 13.5\% were malicious.

\subsubsection{Malware families}

\begin{table}[!ht]
\centering
\small
\begin{tabular}{
p{0.25\textwidth}|
>{\raggedleft\arraybackslash}p{0.1\textwidth}|
>{\raggedleft\arraybackslash}p{0.1\textwidth}|
>{\raggedleft\arraybackslash}p{0.1\textwidth}|
l}
Malware family & \# Train set & \# Test set & Total & Description \\\hline
startsurf   & 66&  273 &  339  & adware \\
fareit  & 33& 222  & 255  & spyware \\
vigram  &  23   & 212  &235  & adware \\
winwrapper  & 78&  8 & 86  & PUA \\
downloadguide  &  15& 59  & 74  & adware \\
gandcrab&  5& 54  & 59  & ransomware\\
emotet  & 12&  46   & 58  & credstealer \\
chapak  & 4 &  37 & 41  &   installer  \\
virut   & 30&  2 & 32  & backdoor \\
installmonster  & 12   & 18  & 30  & installer  \\
noon&  8 & 22  & 30  & spyware \\
gamarue & 11&  18 & 29  & backdoor \\
razy  & 7&   16 & 23  & crypto stealer \\
zeroaccess  & 23   & 0  & 23  & rootkit \\
soft32downloader  &  5  & 22  & 23  & installer  \\
appster  & 7   & 15  & 22  & PUA \\
prepscram  &  1  & 20  & 21  & installer \\
zusy  &  2  & 19  & 21  & spyware \\
darkkomet  & 17   &  1 & 18  & RAT \\
adposhel  &  4  & 14  & 16  & adware \\
swrort  & 13   & 0  & 13  & backdoor\\
slugin  & 13  & 0  & 13  &  installer\\
vobfus  & 11  & 2  & 13  & installer \\
speedingupmypc  &  1  & 11  & 12  & adware \\
relevantknowledge  &  5   &  6 & 11  & adware \\
kuaizip  & 4   &  7 & 11  & PUA \\
bladabindi  & 7   & 4  & 11  & backdoor \\\hline\hline
Other ($\leq$ 10 instances) &  377 & 260 & 602 & - \\\hline
\# Other families ($\leq$ 10 instances) &  184  & 154 & 288  & - \\\hline\hline
Unknown &  333   & 291  &   671 & - \\\hline
\textbf{Total} &  \textbf{1,137}   &  \textbf{1,655} &   \textbf{2,792} & -
\end{tabular}
\caption{
Malware families with more than 10 samples in the dataset. 315 families were represented in the dataset, with 27 having being represented more than 10 times. Basic description provided which does not cover the wide range of behaviours carried out by some malware families but is intended to indicate the range of behaviours in the top 27 families included in the dataset. PUA = potentially unwanted application, RAT = remote access trojan}
\label{tab:malware_families}
\end{table}

This paper is not concerned with distinguishing particular malware families, but rather with identifying malware in general. However, a dataset consisting of just one malware family would present an unrealistic and easier problem than is found in the real-world. The malware families included in this dataset are reported in Table \ref{tab:malware_families}. The malware family labels are derived from the output of around 60 antivirus engines used by VirusTotal \cite{virusTotal}.

Ascribing family labels to malware is non-trivial since antivirus vendors do not follow standardised naming conventions and many malware families have multiple aliases. Sebasti{\'a}n et al. \cite{sebastian2016avclass} have developed an open source tool, AVClass, to extract meaningful labels and correlate aliases between different antivirus outputs. AVClass was used to label the malware in this dataset. Sometimes there is no consensus amongst the antivirus' output or the sample is not recognised as a member of an existing family. AVClass also excludes malware that belongs to very broad classes of malware (e.g. ``agent", ``eldorado", ``artemis") as these are likely to comprise a wide range of behaviours and so may be applied as a default label in cases for which antivirus engines are unsure. In the dataset established in this research, 2,121 of the 2,792 samples were assigned to a malware family. Table \ref{tab:malware_families} gives the number of samples in each family for which there were more than 10 instances found in the dataset. 315 families were detected overall, with 27 families being represented more than 10 times. These better-represented families persist in the train and test sets, but the other families have little overlap. 104 of the 154 other families seen in the test set are not identified by AVClass as being in the training set.

\subsubsection{Malicious vs. Benign Behaviour}

Statistical inspection of the training set reveals that benign applications have fewer sub-processes than malicious processes, with 1.17 processes in the average benign process tree and 2.33 processes in the average malicious process tree. Malware was also more likely to spawn processes outside of the process tree of the root process, often using the names of legitimate Windows processes. In some cases malware launches legitimate applications, such as Microsoft Excel in order to carry out a macro-based exploit. Although Excel is not a malicious application in itself, it is malicious in this context, which is why malicious labels are assigned if a malware sample has caused that process to come into being. It is therefore possible to argue that some processes launched by malware are not malicious, because they do not individually cause harm to the endpoint or user, but without the malware they would not be running and so can be considered at least undesirable even if only in the interests of conserving computational resources.

\subsubsection{Train-Test Split}

The dataset is split in half with the malicious samples in the test set coming from the more recent VirusShare repository, and those in the training set from the earlier repository. This is to increase the chances of simulating a real deployment scenario in which the malware tested contain new functionality by comparison with those in the training set. 

Ideally the benignware should also be split by date across the training and test set, however it is not a trivial task to calculate the date at which benignware was compiled. It is possible to extract the compile time from PE header, but it is possible for the PE author to manually input this date which had clearly happened in some instances where the compile date was 1970-01-01 or in one instance 1970-01-16. In the latter case (1970-01-16), the file is first mentioned online in 2016, perhaps indicating a typographic error \cite{2016benignware}. Using internet sources such as VirusTotal \cite{virusTotal} can give an indication when software was first seen but if the file is not very suspicious i.e. from a reputable source, it may not have been uploaded until years after it was first seen ``in the wild'. Due to the difficulty in dating benignware in the dataset collected for this research, samples were assigned to the training or test set randomly. 

For training, an equal number of benign and malicious processes are selected, so that the model does not bias towards one class. 10\% of these are held out for validation. In most ML model evaluations, the validation set would be drawn from the same distribution as the test set. However, because it is important not to leak any information about the malware in the test set, since it is split by date, the validation set here is drawn from the training distribution. 

\subsubsection{Implementation Tools}

Data collection used the Psutil \cite{psutil} Python library to collect machine activity data for running processes and to kill those processes deemed malicious. The RNN and Random Forests were implemented using the Pytorch \cite{paszke2017automatic} and Scikit-Learn \cite{scikit-learn} python libraries respectively. The model runs with high priority and administrator rights to make sure the polling is maintained when compute resources are scarce.

\section{Experimental Results}\label{experiments}\label{sec:4_results}

\subsection{Supervised Learning for Process Killing}\label{sec:supervised}

First we demonstrate the unsuitability of a full-trace supervised learning malware detection model, which achieved more than 96\% detection accuracy in \cite{rhode2018early}. The model used is a gated-recurrent unit recurrent neural network since this algorithm is designed to process time-series data. The hyperparameter configuration of this model was conducted using a random search of hyperparameters (see table~\ref{tab:app_hyp} in the Appendix for details.)  

It is expected that supervised malware detection models will not adapt well to process-killing due to the averaging of loss metrics as described earlier. Initially this is verified by using supervised learning models to kill processes that are deemed malicious. For supervised classification, the model makes a prediction every time a data measurement is taken from a process. This approach is compared with one taking average predictions across all measurements for a process and for a process tree as well as the result of process killing. The models with the highest validation accuracy for classification and killing are compared. 

\begin{figure}[!ht]
\centering
\includegraphics[width=0.9\textwidth]{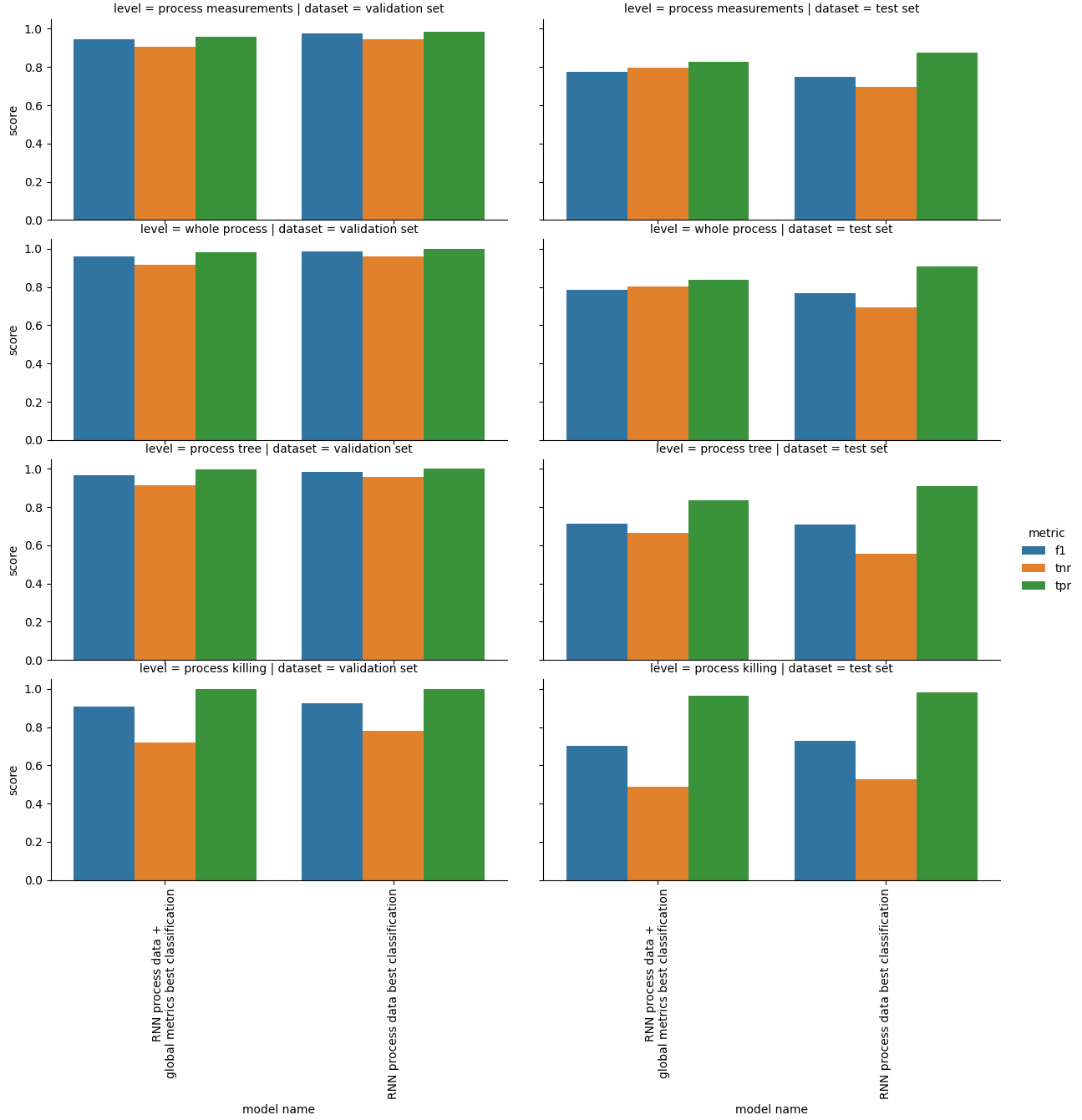}
\caption{F1 scores, true positive rates (TPR) and true negative rates (TNR) for partial-trace detection (process measurements), full-trace detection (whole process), whole application (process tree) and with process level measurements + process killing (process killing) for validation set (left column) and test set (right column)}
\label{fig:4_classify_kill}
\end{figure}

\begin{table}[!ht]
\small
\centering
\begin{tabular}{l|l|l|l|l}
features & metric & classify & dataset & kill \\\hline\hline
          proc. data &     F1 &  \textbf{97.44} &  validation set &       \textbf{91.20} \\
          proc. data &    tnr &  94.72 &  validation set &       85.71 \\
          proc. data &    tpr &  98.64 &  validation set &       95.80 \\\hline
 proc. data + glob. &     F1 &  94.61 &  validation set &       87.69 \\
 proc. data + glob. &    tnr &  90.57 &  validation set &       77.31 \\
 proc. data + glob. &    tpr &  95.93 &  validation set &       95.80 \\\hline\hline
          proc. data &     F1 &  74.91 &        test set &       \textbf{72.63} \\
          proc. data &    tnr &  69.41 &        test set &       59.63 \\
          proc. data &    tpr &  87.52 &        test set &       91.82 \\\hline
 proc. data + glob. &     F1 &  \textbf{77.66} &        test set &       71.83 \\
 proc. data + glob. &    tnr &  79.70 &        test set &       59.63 \\
 proc. data + glob. &    tpr &  82.91 &        test set &       90.24 \\
\end{tabular}
\caption{F1-score, true positive rate (TPR) and true negative rates (TNR) (all * 100) on test and validation sets for classification and process killing}
\label{tab:4_classify_kill}
\end{table}

\begin{figure}
    \centering
    \includegraphics[width=0.9\textwidth]{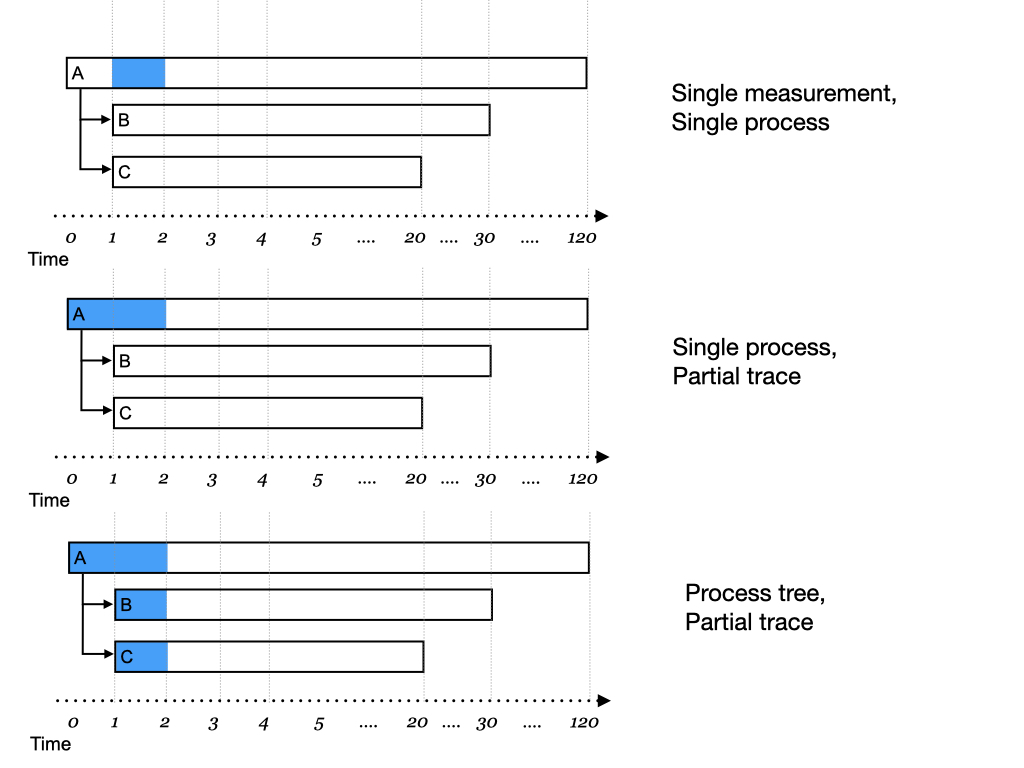}
    \caption{3 levels of data collection: Each measurement, each process, each process tree}
    \label{fig:4_proc_levels}
\end{figure}

Figure \ref{fig:4_classify_kill} illustrates the difference in validation set and test set F1-score, true positive rate and false positive rate for these 4 levels of classification: each measurement, each process, each process tree, and finally showing process killing; see Figures~\ref{fig:4_proc_levels} for diagrammatic representation of these first 3 levels. Table~\ref{tab:4_classify_kill} reports the F1, TPR and TNR for classification (each measurement of each process) and for process killing. 

The highest F1-score on the validation set is achieved by an RNN using process data only. When process killing is applied there is a drop of less than 5 percentage points in the F1-score but more than 15 percentage points are lost from the TNR. 

On the unseen test set the highest F1-score is achieved by an RNN using process data + global metrics, but the improvement over the process data + total number of processes is negligible. Overall is there is a reduction in F1-score from (97.44, 94.61) to (74.91, 77.66), highlighting the initial challenge of learning to classifying individual processes rather than entire applications, especially when accounting for concept drift. Despite the low accuracy, these initial results indicate that the model is discriminating some of the samples correctly and may form a baseline from which to improve.

The test set TNR and TPR for classification on the best-performing model (process data only) are 79.70 and 82.91 respectively, but when process killing is applied, though the F1-score drops by 10 percentage points, the TNR and TPR move in opposite directions with the TNR falling to 59.63 and TPR increasing to 90.24. This is not surprising since a single malicious classification results in a process being classed as malicious. This is true for the best-performing models using either of the two feature sets (see Fig.\ref{fig:4_classify_kill} above).

\subsection{Accuracy vs. Resource consumption}\label{sec:acc_vs_resource}

Previous work on real-time detection has highlighted the requirement for a lightweight model (speed and computational resources). The previous paper, RNNs were the best performing algorithm in classifying malware/benignware but RNNs have many parameters and therefore may consume significant RAM and/or CPU, they also require preprocessing of the data to scale the values, which other ML algorithms such as tree-based algorithms do not. Whilst RAM and CPU should be minimised, taking model accuracy into account, inference duration is also an important metric.

Though the models in this paper have not been coded for performance and use common python libraries, comparing these metrics helps to decide whether certain models are vastly preferable to others with respect to computational resource consumption. The PyRAPL library \cite{pyrapl} is used measure the CPU, RAM and duration used by each model. This library uses Intel processor `Running Average Power Limit' (RAPL) metrics. Only the data pre-processing and inference is measured as training may be conducted centrally in a resource-rich environment. Batch sizes of 1, 10, 100 and 1000 samples are tested with 26 and 37 features respectively since there are 26 process-level features and 37 when global metrics are included. Each model is run 100 times for each of the different batch sizes.

For the RNN a `large' and a `small' model are included. The large models have the highest number of parameters tested in the random search (981 hidden neurons, 3 hidden layers, sequence length of 17) and the smallest (41 neurons, 1 hidden layer, sequence length of 13). These two RNN configurations are compared against other machine learning models which have been used for malware detection: Multi-Layer Perceptron (feed-forward neural network), Support Vector Machine, Naive Bayes Classifier, Decision Tree Classifier, Gradient Boosted Decision Tree Classifier (GBDTs), Random Forest and AdaBoost. 

\begin{table}[!htbp]
\centering\footnotesize
\begin{tabular}{lR{1cm}rR{1.7cm}R{1.7cm}rR{1.2cm}rR{1.2cm}}
    \hline
    model & n features & avg. cpu (\(\mu J\)) &  avg. dram (W) &  avg. duration (\(\mu s\)) &val F1 & kill val F1 &   test F1 &  kill test F1 \\
    \hline
    AdaBoost &   26 &  127967.84 &       7981.51 &          6595.37 &   88.35 &        74.36 &    77.19 &         60.09 \\
 AdaBoost &   37 &  125041.20 &       7142.93 &          6469.16 &   89.63 &        76.07 &    80.10 &         60.14 \\
       DT &   26 &    3905.63 &        202.65 &           128.02 &   97.39 &        88.48 &    66.44 &         62.95 \\
       DT &   37 &    \textbf{2113.67} &        \textbf{134.29} &  \textbf{106.65} &   96.32 &        83.57 &    79.61 &         62.50 \\
     GBDT &   26 &    8788.41 &        338.78 &           349.31 &   92.27 &        78.26 &    82.47 &         63.33 \\
     GBDT &   37 &   11005.80 &        486.46 &           329.45 &   93.13 &        80.26 &    84.94 &         63.46 \\
      MLP &   26 &   11044.88 &        645.14 &           461.04 &   82.84 &        70.18 &    41.62 &         57.65 \\
      MLP &   37 &   12932.09 &        628.64 &           555.42 &   73.00 &        67.63 &    57.66 &         57.26 \\
       NB &   26 &    6947.67 &        297.87 &           185.73 &   75.80 &        67.42 &    62.90 &         56.11 \\
       NB &   37 &    5187.96 &        258.80 &           177.37 &   75.58 &        67.61 &    61.88 &         55.33 \\
       RF &   26 &  238621.20 &      11052.84 &          8997.31 &   97.12 &        \textbf{92.97} &    71.58 &         75.97 \\
       RF &   37 &  236598.44 &       9967.63 &          8879.97 &   96.57 &        91.05 &    \textbf{85.55} &         \textbf{77.85} \\
      RNN &   26 &  887664.31 &      48885.96 &         27869.30 &   \textbf{97.44} &        90.70 &    74.91 &         73.08 \\
      RNN &   37 &  312108.07 &      17120.90 &         10414.58 &   94.61 &        87.31 &    77.66 &         71.95 \\
      SVM &   26 &          6630490.84 &     464082.07 &        282026.57 &   78.34 &        67.04 &    68.16 &         56.91 \\
      SVM &   37 &          7792179.78 &     730786.06 &        429081.31 &   64.89 &        65.68 &    61.39 &         56.25 \\
    \hline
    \end{tabular}
\caption{Average resource consumption over 100 iterations for a batch size of 100 vs. F1-scores on validation and test set for classification and process killing across 14 models. 26 features = process-level only, 37 features = machine and process level features.}
\label{tab:4_algo_comparison}
\end{table}

Table \ref{tab:4_algo_comparison} reports the computational resource consumption and accuracy metrics together. Decision tree with 38 features is the lowest cost to run, RNN performs best at supervised learning classification on the validation set but only just outperforms the decision tree with 26 features, which is the best performing model at process killing on the validation set at 92.97 F1-score. The highest F1-score for process killing uses a Random Forest with 37 features, scoring 77.85 F1, which is 2 percentage points higher than the RF with 26 features (75.97). The models all perform at least 10 percentage points better on the validation set indicating the importance of taking concept drift into account when validating models. 

\subsection{How to solve a problem like process killing?}

From the results above, it is clear that supervised learning models see a significant drop in classification accuracy when processes are killed as the result of a malicious label. This confirmation of the initial hypothesis presented here justifies the need to examine alternative methods. In the interests of future work and negative result reporting this paper reports all of the methods attempted and finds that simple statistical manipulations on the supervised learning models perform better than using alternative training methods. This section briefly describes the logic of each method and provides a textual summary of the results with a formulae where appropriate. This is followed by a table of the numerical results for each method. In the following section let \(P\) be a set of processes \(\{p_{0},p_{1}... p_{P}\}\) in a process tree, let \(t*\) be the time at which a prediction is made and let \(\hat{y}_{i} \) be the prediction for process \(i\) at time t* where a prediction equal to or greater than 1 classifies malware.

\subsubsection{a) Mean predictions}
\textit{Reasoning: Taking the average prediction across the whole process will smooth out those process killing results} 

\textbf{Not tested} This was not attempted for two reasons: (1) Taking the mean at the end of the process means the damage is done (2) This method can easily be manipulated by an attacker: 50 seconds of injected  benign activity required 50 seconds of malicious activity to achieve a true positive 

\[\hat{y}_{i} = \frac{1}{t*}\sum_{t=0}^{t*}{\hat{y_{i}^{t}}}\]

\subsubsection{b) Rolling mean predictions}
\textit{Reasoning: Taking the average over a few measurements will eliminate those false positives that are caused by a single false positive over a subset of the execution trace. Window sizes of 2 to 5 are tested. Let \(w\) be the window size:}

\[\hat{y}_{i} = \frac{1}{w}\sum_{t=t* - w}^{t*}{\hat{y_{i}^{t}}}\]

\textbf{Summary of results:} A small but unilateral increase in F1-Score using a rolling window over 2 measurements on the validation set. Using a rolling window of size 2 on the test-set saw a 10 to 20 percentage point increase in true negative rate (to a maximum of 80.77) with 3 percentage points lost from the true positive rate. This was one of the most promising approaches.

\subsubsection{c) Alert threshold}
\textit{Reasoning: Like the rolling mean, single false positives will be eliminated but unlike the rolling mean, the alerts are cumulative over the entire trace such that a single alert at the start and 30 seconds into the process will cause the process to be killed rather than requiring that both alerts are within a window of time. Between 2 and 5 minimum alerts are tested} 

\[\hat{y}_{i} = w - \sum_{t=0}^{t*}{\hat{y_{i}^{t}}}\]

\textbf{Summary of results:} Again a small increase across all models, with an optimal minimum number of alerts being 2 for maximum F1-score, competitive with the rolling mean approach.

\subsubsection{d) Process-tree averaging}
\textit{Reasoning: the data are labelled at the application level, therefore the average predictions across the process tree should be considered for classification}

\[\hat{y}_{i} = \frac{1}{P}\sum_{p=0}^{P}{\hat{y_{i}^{t*}}}\]

\textbf{Summary of results:} Negligible performance increase on validation and test set data (less than 1 percentage point). This is likely because few samples have more than one process executing simultaneously.

\subsubsection{e) Process-tree training}
\textit{Reasoning: the data are labelled at the application level, therefore the sum of resources of each process tree should be classified at each measurement, not the individual processes}

\textbf{Summary of results:} Somewhat surprisingly there was a slight reduction in classification accuracy when using process tree data. One explanation for this may be that the process tree creates noise around the differentiating characteristics that are visible at the process level.

\subsubsection{DQN}
\textit{Reasoning: Reinforcement learning is designed for state-action space learning. Both pre-training the model with a supervised learning approach and not pre-training the model were tested.}

\textbf{Summary of results:} Poor performance, typically converging to either kill or not kill everything, of the few models which did not converge to a single dominant action, it does not distinguish malware or benignware well, indicating that it may not have learned anything. Reinforcement learning may help the problem of real-time malware detection and process killing but this initial implementation of a DQN did not converge to a better or even competitive solution to supervised learning. Perhaps better formulation of rewards (e.g. damage prevented) would help the agent learn.

\subsubsection{Regression on predicted kill value}
\textit{Reasoning: Though the DQN explores and exploits different state-action pairs and their associated rewards, when the reward from each action is known in the first place and the training set is limited, as it is here, Q-learning can be framed as a regression problem in which the model tries to learn the return (rewards + future rewards), the training is faster and can be used by any regression-capable algorithm. Let \(N\) be the number of current and future child processes for \(p_{i}\) at \(t*\)}

\[(y * 2 - 1) * (1 + N) * (1 + (y * (e^{-t*})))\]

\textbf{Summary of results:} Improved performance on true negative rate, though not perceptible for the highest-scoring F1 models since F1-scores reward true positives more than true negatives, this metric can struggle to reflect a balance between the true positive and true negative rates. The highest true negative rate models are all regression models.

\begin{table}[ht!]\small
    \centering
    \begin{tabular}{p{2.4cm}|p{1.3cm}|l|p{1cm}|r|r|r|r|r|r}
                       & & & \multicolumn{3}{|c}{val} & \multicolumn{3}{|c}{test} \\
        methodology & best dataset & model  & n features &     F1 &     tnr &    tpr &     F1 &    tnr &    tpr \\\hline
        Supervised learning & val & RF &  26 &   92.37 &   87.39 &  96.64 &  74.57 &  62.71 &  92.95 \\
                            & test & RF &  37 &  89.68 &   83.19 &  94.96 &  76.43 &  67.19 &  92.52 \\
                            
        Rolling mean         & val &  RF (min: 2) &  26 &  \textbf{93.22} &   94.12 &  92.44 &  78.26 &  73.83 &  89.76 \\
                             & test  &  RF (min: 2) &  37 &  92.70 &   94.96 &  90.76 &  80.77 &  78.88 &  89.38 \\
        
        Alert threshold     & val &  DT (min: 2) &  26 &  92.17 &   95.80 &  89.08 &  73.43 &  67.44 &  86.56 \\
                        & test & RF (min: 2) &  37 &  91.30 &   94.96 &  88.24 &  \textbf{81.50} &  81.53 &  87.97 \\
        
        Process tree averaging & val & RF &  26 &  92.74 &   88.24 &  96.64 &  74.79 &  64.04 &  92.20 \\
                            & test & RF &  37 &  90.48 &   84.03 &  95.80 &  76.34 &  67.66 &  91.92 \\
        
        Process tree training & val &  RF  &  26 &  90.35 &   82.58 &  98.32 &  74.20 &  52.44 &  92.74 \\
                                & test & RF &  26 &  90.35 &   82.58 &  98.32 &  74.20 &  52.44 &  92.74 \\
        
        Q-learning & val & DQN & 26 &  51.71 &   72.27 &  44.54 &  27.74 &  55.50 &  26.94 \\
                    & test & DQN & 26 &  51.71 &   72.27 &  44.54 &  27.74 &  55.50 &  26.94 \\
        
        Regression  & val &  RF &  26 &  91.94 &   87.39 &  95.80 &  74.77 &  66.05 &  90.35 \\
                    & test & RF &  26 &  91.94 &   87.39 &  95.80 &  74.77 &  66.05 &  90.35 \\

    \end{tabular}
    \caption{Summary of best process killing models by model training methodology. F1, TNR and TPR for validation and test datasets. Full results in Appendix Tables \ref{app_tab:paper4_process_killing}, \ref{app_tab:chapter4_process_killing_2} and  \ref{app_tab:chapter4_process_killing_3}}
    \label{tab:paper4_process_killing}
\end{table}

Table \ref{tab:paper4_process_killing} lists the F1, TPR and TNR on the validation and test set for each of the methods described above. The best-performing model on the test and validation sets are reported and the full results can be found in Appendix Table ~\ref{app_tab:paper4_process_killing}. Small improvements are made by some models on the validation F1-score but the test set F1-score improves by 4 percentage points in the best instance.

In most cases, the models with the highest F1-score on the validation and test sets are not the same. The highest F1-score is 81.50 from an RF using a minimum alert threshold of 2 and both process-level and global process metrics. 

\subsection{Further experiment: Favouring high TNR}

Though the proposed model is motivated by the desire to prevent malware from executing, the best TNR reported amongst the models above is 81.50\%. 20\% of benign processes being killed would not be acceptable to a user. Whilst this research is a novel attempt at very early-stage real-time malware detection and process killing, one might consider the usability and prefer a model with a very high TNR, even if this is at the expense of the TPR. 

\begin{table}[ht!]
    \centering\small
    \begin{tabular}{l|l|l|r|r|r|r|r|r|r}
                       & & & \multicolumn{3}{c}{val} & \multicolumn{3}{c}{test} \\
        methodology & model  & n features &     F1 &     tnr &    tpr &     F1 &    tnr &    tpr \\\hline
        Regression &   AdaBoost & 26            & 56.63 & 100.00 & 39.50 & 15.06 & 97.92 & 8.40 \\
        Regression + 4 alerts & GBDT & 26 & 85.91 & 95.80 & 77.31 & 68.50 & 94.98 & 56.04
    \end{tabular}
    \caption{Two models' F1-score, TNR, TPR for the validation and test set scoring the highest TNR on the validation and test sets.}
    \label{tab:4_highTNR}
\end{table}

Considering this, the AdaBoost regression algorithm achieves a 100\% TNR with a 39.50\% TPR on the validation set. The high FNR is retained in the test set standing at 97.92\% but the TPR drops even further to just 8.40\%. The GBDT also using regression to estimate the value of process killing, and coupled with a minimum of 4 alerts performs well on the test set but does not stand out in the validation set see Table~\ref{tab:4_highTNR}. 

Though less than 10\% of the test set malicious processes are killed by the AdaBoost regressor, this model may be the most viable despite the low TPR. Future work may examine the precise behaviour and harm caused by malware that is/is not detected. To summarise results, the most-detected families were Ekstak (180), Mikey (80), Prepscram (53 processes) and Zusy (49 processes) of 745 total samples. 

\subsection{Measuring damage prevention in real time}\label{sec:4_ransomware}

Though a high percentage of processes are correctly identified as malicious by the best performing model (RF with 2 alerts and 37 features); it may be that the model detects the malware after it has already caused damage to the endpoint. Therefore, instead of looking at the time at which the malware is correctly detected, a live test was carried out with ransomware to measure the percentage of files corrupted with and without the process killing model working. This real time test also assesses whether malware can indeed be detected in the early stages of execution or whether the data recording, model inference and process killing is too slow in practice to prevent damage.

Ransomware is the broad term given to malware that prevents access to user data (often by encrypting files) and holds the means for restoring the data (usually a decryption key) from the user until a ransom is paid. It is possible to quantify the damage caused by ransomware using the proportion of modified files as Scaife et al. \cite{scaife2016cryptolock} have done in developing a real-time ransomware (only) detection system. The damage of some malware types are more difficult to quantify owing to their dependence on factors outside the control of the malware. For example the damage caused by spyware will depend on what information it is able to obtain so it is difficult to quantify the benefit of killing spyware 5 seconds after execution compared with 5 minutes into execution. Ransomware offers a clear metric for the benefits of early detection and process killing.

\begin{table}[ht!]
    \centering
    \begin{tabular}{l|l|r|r|r|r|r|r}
         & & \multicolumn{3}{c}{val} & \multicolumn{3}{c}{test} \\
        model  & n features &     F1 &     tnr &    tpr &     F1 &    tnr &    tpr \\\hline
        RF (alerts: 2) &  37 &  91.30 &   94.96 &  88.24 &  \textbf{81.50} &  81.53 &  87.97 \\
        DT (rolling mean: 2) & 26 &  \textbf{93.16} &   94.96 &  91.60 &  73.82 &  66.19 &  88.40 \\
    \end{tabular}
    \caption{Random Forest and Decision Tree each with a minimum requirement of two alerts (`malicious classifications') to kill a process. F1, TNR and TPR reported on validation and test set.}
    \label{tab:4_RFvsDT}
\end{table}

Although the RF with with a minimum of 2 alerts using both process and global data gave the highest F1-score on the test set (81.50), earlier experiments showed that RFs are not one of the most computationally efficient models by comparison with those tested. Therefore a decision tree is trained on process-only data (26 features) in case the time-to-classification is important for damage reduction despite the lower F1-score. For this reason the decision tree model is used in this test. The DT also has a very slightly higher TPR (see Table \ref{tab:4_RFvsDT}) so a higher damage prevention rate may be partially due to the model itself rather than just the fewer features being collected and model classification speed.

22 fast-acting ransomware files were identified from a separate VirusShare \cite{VirusShare} repository which \textit{(i)} do not require internet connection and \textit{(ii)} begin encrypting files within the first few second of execution. The former condition is set because the malicious server may no longer exist and for safety it is not desirable to connect to it if it does. These are the types of malware that if the proposed model could block, would save significant damage to the user in a time-frame that it would be difficult for a human to react to.

The 22 samples were executed for 30 seconds each without the process killing model and the number of files modified was recorded. The process was repeated with 4 process killing models: DT with min. 2 alerts and 26 features, RF with min. 2 alerts and 37 features, AdaBoost regressor with 26 features and GDBT regressor with min. 4 alerts and 26 features.

It was necessary to run the killing model with administrator privileges and to write an exception for the Cuckoo sandbox agent process which enables the host machine to read data from the guest machine since the models killed this process. The need for this exception highlights that there are benign applications with malicious-like behaviours perhaps especially those used for networking and security. 

\begin{figure}[ht!]
    \centering
    \includegraphics[width=1\textwidth]{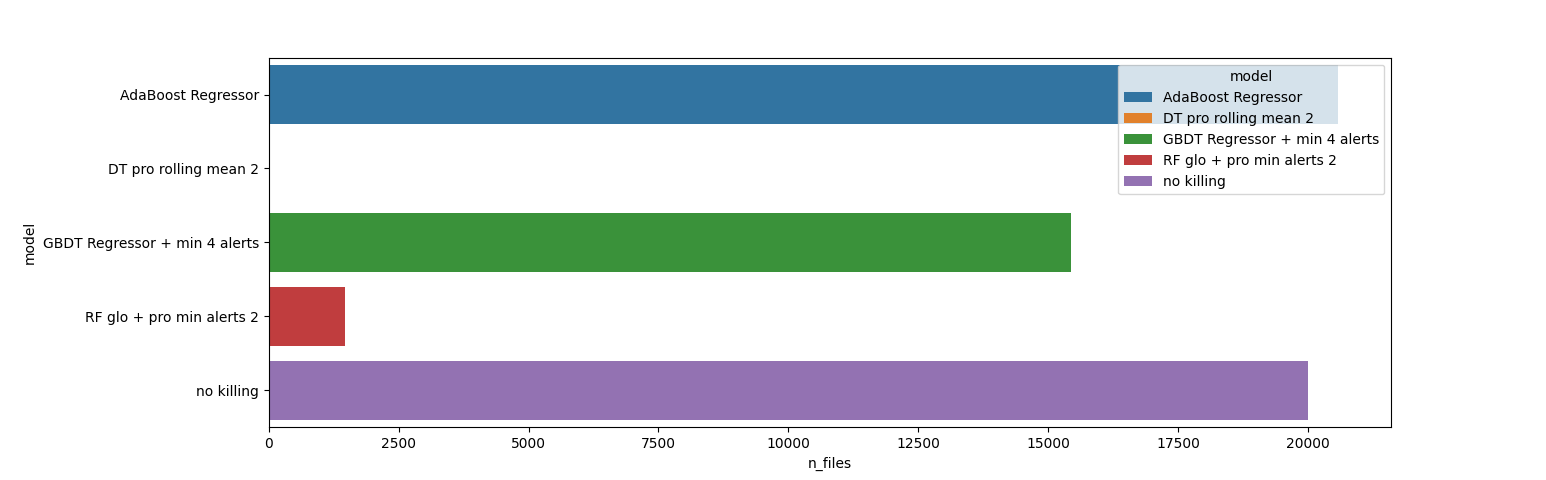}
    \caption{Total number of files corrupted by ransomware with no process killing and with three process killing models within the first 30 seconds of execution.}
    \label{fig:4_realkill}
\end{figure}

\begin{table}[ht!]
    \centering
    \begin{tabular}{l|R{2cm}|R{2cm}|R{2cm}|R{2cm}}
                             Model &  Files damaged & Damage reduction & Detection rate (Ransomware TPR) & Test set TPR \\\hline
                        no killing &    19,997 & - & - & -\\

             DT pro rolling mean 2 &        3  & 99.98\% & 100.00 & 88.40\\
            RF glo + pro min alerts 2 &     1,464 & 92.68\% & 100.00 & 87.97\\
     GBDT regressor + min 4 alerts &    15,432 & 22.83\% & 22.07 & 56.04\\
                     AdaBoost regressor &    20,578 & 0.00\% & 9.09 & 8.83\\

    \end{tabular}
    \caption{Total number of files corrupted by ransomware with no process killing and with three process killing models within the first 30 seconds of execution. Damage reduction is the percentage of files spared when no killing is implemented }
    \label{tab:4_realkill}
\end{table}

Figure \ref{fig:4_realkill} and Table \ref{tab:4_realkill} give the total number of corrupted files across the 22 samples. The damage prevention column is a proxy metric denoting how many files were not corrupted using a given process killing model by comparison with no model being in place. The 22 samples on average each corrupt 910 files within 30 seconds. 

The DT model almost entirely eliminates any file corruption with only three being corrupted. The RF saves 92.68\% of files. The ordinal ranking of `damage prevention' is the same as the TPR on the test set but the relationship is not proportional. The same ordinal relationship indicates that the simulated impact of process killing on the collected test set was perhaps a reasonable approximation of measuring at least fast-acting ransomware damage, despite the TPR test set metrics being based on other malware families too.

The DT demonstrates that this architecture is capable of preventing damage, but the TNR on the test set of the DT model is so low (66.19) that this model cannot be preferred to the RF (81.53 TNR), which still prevents over 90\% of file damage. 

The GBDT prevents some damage, and detects a comparable number of ransomware samples (1 in 5). The AdaBoost regressor detected 2 ransomware samples of the 22, and in these two cases more than 64\% and 45\% of files were saved respectively; perhaps with more execution time the files would be detected but the key benefit of process killing is to stop damaging software like these ransomware samples and this algorithm actually saw more files encrypted than when no killing model was used; this is because there will be a slight variance in the ransomware behaviour and execution time each time it runs. The Random Forest is the most plausible model, balancing damage prevention and TNR, however the delay in classification may be a result of the requirement to collect more features and/or the real-time of the model itself. 

\section{Discussion: Measuring execution time in a live environment}\label{sec:4_discussion}

Though algorithm execution duration was measured above, due to batch processing used by the models, the number of processes being classified can be increased by an order of magnitude with a negligible impact on execution time. The data collection and process killing both have \(O(n)\) complexity; where \(n\) is the number of processes therefore it is expected that the number of processes to impact classification time. The RF with statistical filters has complexity \(O(nps)\) where \(p\) are the number of trees in the forest and \(s\) is the number alerts considered by the filter, efficient library implementations of matrix operations mean that the execution time does not scale linearly with \(n\) for the RF inference. Given this, a further experiment was carried out with the RF to measure in a live environment how long the data collection, model inference, and process killing takes as the number of processes increases. This was tested by executing more than 1000 processes in the virtual machine whilst the process killing RF runs.

\begin{figure}
    \centering
    \includegraphics[width=1\textwidth]{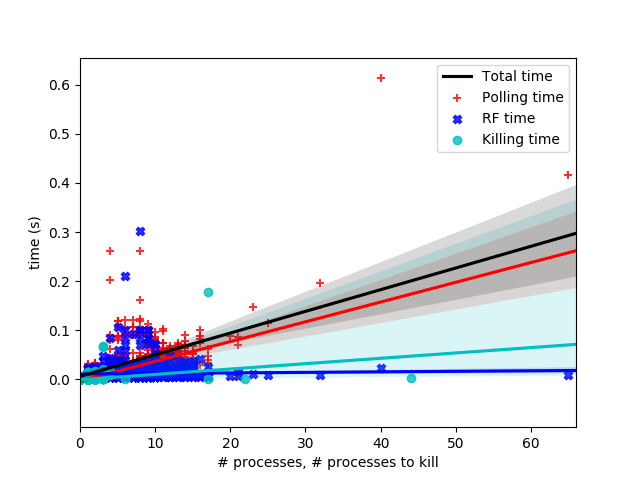}
    \caption{Mean time to collect data, analyse data with Random Forest, and kill varying numbers of processes}
    \label{fig:performance}
\end{figure}

Some processes demand more computational resources than others, and some malware in our test set locked pages in memory\cite{lock_memory} which prevented the model from having sufficient resources to collect data, leading to tens of seconds during which no data was captured and many processes were launched, with better software engineering practices the model may be more robust against this kind of malicious activity. 

These differences in behaviour can cause the evaluation time to lag as demonstrated by the outlier points visible in Figure~\ref{fig:performance}. The data shows a broadly linear positive correlation between the number of processes (being monitored or killed) and the time taken for the data collection and process killing; this confirms the hypothesis that more processes equates to slower processing time. The slowest total processing time was 0.81 seconds (seen with both 17 and 40 simultaneous processes running) but the mean processing time is just under 0.3 seconds with 65 simultaneous processes, fitting comfortably within the 1-second goal time. Additional code optimisation could  greatly improve on these initial results which indicate that the processing, even using standard libraries and a high level programming language, can execute reasonably quickly.  

\section{Implications and Analysis}\label{sec:implications}

The experiments in this paper address a largely unexplored area of malware detection, by comparison with post-trace classification. Real-time processing and response has a number of benefits, outlined above and the results presented here give tentative indications of the advantages and challenges of such an approach.

The initial experiments (Section \ref{sec:supervised}, demonstrate that a high-accuracy RNN (as used in \cite{rhode2018early}) does not maintain high-accuracy when used in real-time with an automated response to classify individual processes rather than full application traces, since a single false positive classification of sequential data cannot be outweighed by later correct predictions.

The next set of experiments (Section \ref{sec:supervised}) showed that whilst the RNN achieves one of the highest classification accuracies of a set of algorithms tested, it is not one of the best in terms of computational resource consumption or latency. However, a clear best-algorithm was not evident either since the low-resource consuming algorithms (like decision tree) did not always achieve high accuracy. Furthermore, all of the supervised learning algorithms were clearly unsuited to process killing with the highest F1 score from any algorithm being 77.85 on the test set compared with 85.55 for process-level classification alone. This 85.55 F1 score is lower than is seen in many dynamic malware detection research publications that use full-application behavioural traces, indicating the challenges of classification at the process level, where malware and benignware may share functionality.

Attempting to improve detection accuracy, three approaches were tested: statistical filtering, reinforcement learning and a regression model estimating the utility (q-value) of killing a process. Statistical filters using rolling mean or alert thresholds were the only approach to improve on the supervised learning model F1 score. Reinforcement learning tended to kill processes too early and therefore not explore enough scenarios (and thus receive the requisite reinforcement) to allow benign processes to continue, this does not mean that future models could not improve upon this result. This may be supported by the success of the regression models in maintaining a high true-negative rate, given that these models ascribed the similar utility to killing processes as the reinforcement learning models.

The accuracy metrics tested thus far simply indicate whether a process was ever killed, but do not address whether damage was actually prevented by process killing. If damage was not prevented, there is little point to process killing and a database of alerts for analysis would be a better solution since the risk of killing benignware is eliminated. This is why the final set of experiments in Section~\ref{sec:4_ransomware} were conducted to test the detection models in real-time and see if damage could be prevented by looking at the number file corrupted by ransomware before and after infection. Here we found that is was possible to prevent 92\% of files from being encrypted whilst maintaining a true negative rate of 82\%. This result does not indicate that the system is ready for real-world deployment but that perhaps further model analysis perhaps including anomaly detection could raise the true negative rate to a usable point. This work also demonstrates the damage that certain malware can carry out in a short space of time and reinforces the need for further research in this area, since previous work has either focused solely on ransomware \cite{scaife2016cryptolock} or waited minutes to being classification \cite{fastSlow2017}, by which time it is too late.

\section{Future Work}\label{sec:4_future_work}

Real-time attack detection has wider applications than endpoint detection, as Alazab et al. \cite{alazab2021federated} argue, Internet of Things networks in particular could benefit from real-time attack detection using heterogeneous data feed from different sensors combined using federated learning approaches.

However, some challenges remain to be solved; behavioural malware analysis research using machine learning regularly reports $>$95\% classification accuracy. Though useful for analysts, behavioural detection should be deployed as part of endpoint defensive systems to leverage the full benefits of a detection model. Dynamic analysis is not typically used for endpoint protection, perhaps because it takes too long in data collection to deliver the quick verdicts required for good user experience. Real-time detection on the endpoint allows for observation of the full trace without the user having to wait. However, real-time detection also introduces the risk that malware will cause damage to the endpoint. This risk requires that processes detected as malicious are automatically killed as early as possible to avoid harm. 

There are some key challenges to implementation, which have been outlined in this paper: 

\begin{itemize}
    \item The need for signal separation drives the use of individual processes and only partial traces can be used. 
    \item The significant drop in accuracy on the unseen test set, even without process killing demonstrates that additional features may be necessary to improve detection accuracy.
    \item With the introduction of process killing, the poor performance of the models on either benignware classification (RF min 2 alerts: TNR 81\% with an 88\% TPR on the test set) or on malware classification (GBDT regressor min 4 alerts: 56\% TPR with a 94\% TNR on the test set) means that considerable further work is needed before very early stage real-time detection can be considered for real-world use.
    \item Real-time detection using full execution traces of processes however, may be viable. This is useful to handle VM-aware malware which may only reveal its true behaviour in the target environment. Although the more complex approach using DQNs algorithms did not outperform the supervised models with some additional statistical thresholds, the regression models had better performance in correctly classifying benignware. Reinforcement learning could still be useful for real-time detection and automated cyber defence models, but the DQN in these experiments did not perform well.
    \item  Despite the theoretical unsuitability of supervised learning models to state-action problems, these experiments demonstrate how powerful supervised learning can be for classification problems, even if the problem is not quite the one that the model is attempting to solve.
    \item Future work may require a more comprehensive manual labelling effort at the process level and perhaps labelling sub-sections of processes as malicious or benign.
\end{itemize}

An additional consideration for real-time detection with automated actions is whether this introduces an additional denial-of-service vector using process injection for example to trigger process killing. This may also however indicate that an attacker is present and therefore aid the user.

\section{Conclusions}\label{sec:4_concl}

This paper has built on previous work in real-time detection to address some of the key challenges: signal separation, detection with partial execution traces and computational resource consumption with a focus on preventing harm to the user, since real-time detection introduces this risk. 

Behavioural malware detection using virtual machines is a well-established research field yielding high detection accuracy in recent literature \cite{Huang2016MtNet, das2016semantics, Ijaz2019StaticAndDynamic, rhode2018early}. However, as is shown here, fixed-time execution in a sandbox may not reveal malicious functionality. Real-time malware analysis addresses this issue but risks executing malware on the endpoint and requires detection to take place at the process level, which is more challenging as the definition of a malicious process can be unclear. These two reasons may account for the limited literature on real-time detection. Looking forward real-time detection may become more popular if static data manipulation and VM-evasion continue to be used and the costs of malicious execution continue to rise. Real-time detection does not need to be an alternative to these approaches, but could hold complementary value as part of a defence-in-depth endpoint security. 

To the best of our knowledge previous real-time detection work has used up to 5 simultaneous applications, whereas users may use far more. This paper has demonstrated that up to 35 simultaneous applications (and nearly 100 simultaneous processes) can be constantly monitored, where previous work \cite{fastSlow2017} had tested a maximum of 5. Moreover, these results demonstrated that data collection presented a greater limiting factor than machine learning algorithms which can easily process 1000 samples with negligible impact on performance. This result is not too surprising since batch processing allows algorithms to achieve O(1) complexity by comparison with O(n) for data collection. 

Automatic actions are necessary in response to a detection if the goal is to prevent harm. Otherwise this is equivalent to letting the malware fully execute and simply monitor it's behaviour since human response times are unlikely to be quick enough for fast-acting malware. From a user perspective the question is not `What percentage of malware was executed?' or `Was the malware detected in 5 or 10 minutes?' but `How much damage as been done?'. 

This paper found that using simple statistical filters on top of supervised learning models it was possible to prevent 92\% of files from being corrupted by fast-acting ransomware thus reducing the requirements on the user or organisation to remediate the damage, since it was prevented in the first instance (the rest of the attack vector would remain a concern). 

This approach does not achieve the detection accuracies of state of the art offline behavioural analysis models but, as stated in the introduction, these models typically use the full post-execution trace of malicious behaviour. Delaying classification until post-execution negates the principal advantages of real-time detection. However, the proposed model presents an initial step towards a fully automated endpoint protection model which becomes increasingly necessary as adversaries become more and more motivated to evade offline automated detection tools. 

\bibliographystyle{ieeetr}
\bibliography{main}

\section*{Appendix}
\appendix
 
\begin{table}[htpb]
\small
\centering
\begin{tabular}{p{0.2\textwidth}|p{0.2\textwidth}|p{0.2\textwidth}|p{0.2\textwidth}}
 & possible values & process level data & process level data + global metrics \\\hline
Hyperparamter &&& \\\hline
hidden neurons & 8 - 1024 & 253  & 193 \\
depth & [1, 2, 3] & 2  & 2 \\
batch size & [64, 128, 256] & 128  & 256 \\
epochs & 1 - 200 & 89  & 161 \\
dropout rate & [0, 0.1, 0.2, 0.3, 0.4, 0.5] & 0.1 & 0.1 \\
window size & 2 - 30 & 16 & 6 \\
\hline
loss function & \multicolumn{3}{c}{binary cross-entropy} \\
weight update rule & \multicolumn{3}{c}{Adam \cite{adam}} \\
recurrent unit & \multicolumn{3}{c}{GRU cell} \\\hline
validation F1-score (*100) & - & \textbf{97.43} & 94.61 \\
\end{tabular}
\caption{Hyperparameter search space and the hyperparameters of the model giving the lowest mean false-positive and false-negative rates}
\label{tab:app_hyp}
\end{table}

\begin{table}[ht]
\tiny
    \centering
    \begin{tabular}{l|r|r|r|r|r|r}
  & \multicolumn{3}{c}{val} & \multicolumn{3}{c}{test} \\
 model &     f1 &     tnr &    tpr &     f1 &    tnr &    tpr \\\hline
 AdaBoostModel\_glo\_pro &  77.58 &   55.46 &  91.60 &  67.04 &  49.80 &  88.67 \\
    AdaBoostModel\_glo\_pro mean process tree &  78.01 &   55.46 &  92.44 &  66.75 &  50.48 &  87.59 \\
  AdaBoostModel\_glo\_pro process tree min alerts: 1 &  78.87 &   55.46 &  94.12 &  62.29 &  34.03 &  90.35 \\
  AdaBoostModel\_glo\_pro process tree min alerts: 2 &  78.87 &   55.46 &  94.12 &  62.29 &  34.03 &  90.35 \\
  AdaBoostModel\_glo\_pro process tree min alerts: 3 &  78.87 &   55.46 &  94.12 &  62.29 &  34.03 &  90.35 \\
  AdaBoostModel\_glo\_pro process tree min alerts: 4 &  78.87 &   55.46 &  94.12 &  62.29 &  34.03 &  90.35 \\
AdaBoostModel\_glo\_pro rolling mean window: 2 &  79.22 &   70.59 &  84.87 &  69.57 &  60.88 &  84.88 \\
AdaBoostModel\_glo\_pro rolling mean window: 3 &  79.37 &   72.27 &  84.03 &  69.59 &  61.53 &  84.39 \\
AdaBoostModel\_glo\_pro rolling mean window: 4 &  80.67 &   80.67 &  80.67 &  68.44 &  67.80 &  77.34 \\
    AdaBoostModel\_glo\_pro sum alerts min: 2 &  80.66 &   78.15 &  82.35 &  69.35 &  66.58 &  79.89 \\
    AdaBoostModel\_glo\_pro sum alerts min: 3 &  81.20 &   83.19 &  79.83 &  67.83 &  70.92 &  73.88 \\
    AdaBoostModel\_glo\_pro sum alerts min: 4 &  80.87 &   84.87 &  78.15 &  65.92 &  73.32 &  69.00 \\
 AdaBoostModel\_pro &  75.34 &   47.06 &  92.44 &  65.64 &  45.79 &  88.89 \\
  AdaBoostModel\_pro mean process tree &  75.86 &   48.74 &  92.44 &  65.74 &  47.83 &  87.59 \\
AdaBoostModel\_pro process tree min alerts: 1 &  75.68 &   45.38 &  94.12 &  60.31 &  26.46 &  91.17 \\
AdaBoostModel\_pro process tree min alerts: 2 &  75.68 &   45.38 &  94.12 &  60.31 &  26.46 &  91.17 \\
AdaBoostModel\_pro process tree min alerts: 3 &  75.68 &   45.38 &  94.12 &  60.31 &  26.46 &  91.17 \\
AdaBoostModel\_pro process tree min alerts: 4 &  75.68 &   45.38 &  94.12 &  60.31 &  26.46 &  91.17 \\
    AdaBoostModel\_pro rolling mean window: 2 &  78.03 &   64.71 &  86.55 &  69.35 &  59.63 &  85.47 \\
    AdaBoostModel\_pro rolling mean window: 3 &  77.99 &   67.23 &  84.87 &  68.91 &  59.20 &  84.99 \\
    AdaBoostModel\_pro rolling mean window: 4 &  80.83 &   79.83 &  81.51 &  69.01 &  66.44 &  79.40 \\
  AdaBoostModel\_pro sum alerts min: 2 &  81.12 &   75.63 &  84.87 &  67.91 &  61.06 &  81.68 \\
  AdaBoostModel\_pro sum alerts min: 3 &  81.17 &   80.67 &  81.51 &  66.64 &  64.97 &  76.42 \\
  AdaBoostModel\_pro sum alerts min: 4 &  79.66 &   80.67 &  78.99 &  64.25 &  68.12 &  70.14 \\
AdaBoostModel\_pro\_tree &  75.08 &   47.73 &  94.96 &  64.12 &  30.58 &  86.60 \\
    AdaBoostRegression\_pro\_process &  56.63 &  100.00 &  39.50 &  15.06 &  97.92 &   8.40 \\
  DTModel\_glo\_pro &  84.53 &   71.43 &  94.12 &  71.41 &  58.19 &  90.62 \\
 DTModel\_glo\_pro mean process tree &  85.93 &   73.95 &  94.96 &  71.49 &  59.48 &  89.70 \\
  DTModel\_glo\_pro process tree min alerts: 1 &  84.64 &   70.59 &  94.96 &  65.59 &  42.42 &  91.27 \\
  DTModel\_glo\_pro process tree min alerts: 2 &  84.64 &   70.59 &  94.96 &  65.56 &  42.34 &  91.27 \\
  DTModel\_glo\_pro process tree min alerts: 3 &  84.64 &   70.59 &  94.96 &  65.56 &  42.34 &  91.27 \\
  DTModel\_glo\_pro process tree min alerts: 4 &  84.64 &   70.59 &  94.96 &  65.56 &  42.34 &  91.27 \\
DTModel\_glo\_pro rolling mean window: 2 &  88.70 &   88.24 &  89.08 &  75.09 &  70.49 &  86.94 \\
DTModel\_glo\_pro rolling mean window: 3 &  87.87 &   87.39 &  88.24 &  74.57 &  69.77 &  86.61 \\
DTModel\_glo\_pro rolling mean window: 4 &  89.08 &   93.28 &  85.71 &  74.04 &  74.29 &  81.63 \\
 DTModel\_glo\_pro sum alerts min: 2 &  89.74 &   91.60 &  88.24 &  75.48 &  72.64 &  85.69 \\
 DTModel\_glo\_pro sum alerts min: 3 &  89.47 &   94.12 &  85.71 &  74.00 &  75.62 &  80.38 \\
 DTModel\_glo\_pro sum alerts min: 4 &  88.39 &   94.96 &  83.19 &  70.19 &  77.30 &  72.63 \\
DTModel\_pro &  89.76 &   82.35 &  95.80 &  71.53 &  56.54 &  92.25 \\
     DTModel\_pro mean process tree &  90.91 &   84.03 &  96.64 &  72.13 &  59.38 &  91.06 \\
DTModel\_pro process tree min alerts: 1 &  90.20 &   82.35 &  96.64 &  64.45 &  37.04 &  92.79 \\
DTModel\_pro process tree min alerts: 2 &  90.20 &   82.35 &  96.64 &  64.42 &  36.97 &  92.79 \\
DTModel\_pro process tree min alerts: 3 &  90.20 &   82.35 &  96.64 &  64.42 &  36.97 &  92.79 \\
DTModel\_pro process tree min alerts: 4 &  90.20 &   82.35 &  96.64 &  64.42 &  36.97 &  92.79 \\
DTModel\_pro rolling mean window: 2 &  93.16 &   94.96 &  91.60 &  73.82 &  66.19 &  88.40 \\
DTModel\_pro rolling mean window: 3 &  91.77 &   94.96 &  89.08 &  73.49 &  66.15 &  87.80 \\
DTModel\_pro rolling mean window: 4 &  90.75 &   95.80 &  86.55 &  72.05 &  69.38 &  82.38 \\
     DTModel\_pro sum alerts min: 2 &  92.17 &   95.80 &  89.08 &  73.43 &  67.44 &  86.56 \\
     DTModel\_pro sum alerts min: 3 &  90.75 &   95.80 &  86.55 &  71.53 &  69.63 &  81.25 \\
     DTModel\_pro sum alerts min: 4 &  89.29 &   95.80 &  84.03 &  67.58 &  70.81 &  73.55 \\
  DTModel\_pro\_tree &  85.93 &   73.48 &  97.48 &  70.40 &  43.02 &  91.57 \\
DTRegression\_pro\_process &  89.06 &   80.67 &  95.80 &  71.62 &  57.98 &  91.22 \\
 GBDTModel\_glo\_pro &  80.44 &   63.87 &  91.60 &  72.62 &  59.73 &  91.71 \\
  GBDTModel\_glo\_pro mean process tree &  80.88 &   63.87 &  92.44 &  72.76 &  60.63 &  91.22 \\
GBDTModel\_glo\_pro process tree min alerts: 1 &  81.75 &   63.87 &  94.12 &  66.32 &  43.28 &  92.14 \\
GBDTModel\_glo\_pro process tree min alerts: 2 &  81.75 &   63.87 &  94.12 &  66.32 &  43.28 &  92.14 \\
GBDTModel\_glo\_pro process tree min alerts: 3 &  81.75 &   63.87 &  94.12 &  66.32 &  43.28 &  92.14 \\
GBDTModel\_glo\_pro process tree min alerts: 4 &  81.75 &   63.87 &  94.12 &  66.32 &  43.28 &  92.14 \\
    GBDTModel\_glo\_pro rolling mean window: 2 &  85.12 &   83.19 &  86.55 &  76.06 &  71.50 &  87.80 \\
    GBDTModel\_glo\_pro rolling mean window: 3 &  84.52 &   84.03 &  84.87 &  75.87 &  71.82 &  87.15 \\
    GBDTModel\_glo\_pro rolling mean window: 4 &  84.12 &   86.55 &  82.35 &  75.25 &  76.69 &  81.57 \\
  GBDTModel\_glo\_pro sum alerts min: 2 &  84.87 &   84.87 &  84.87 &  76.46 &  75.65 &  84.66 \\
  GBDTModel\_glo\_pro sum alerts min: 3 &  85.22 &   89.08 &  82.35 &  74.12 &  78.77 &  77.78 \\
  GBDTModel\_glo\_pro sum alerts min: 4 &  84.44 &   90.76 &  79.83 &  71.99 &  81.10 &  72.30 \\
     GBDTModel\_pro &  80.73 &   62.18 &  93.28 &  71.31 &  58.09 &  90.51 \\
  GBDTModel\_pro mean process tree &  82.05 &   64.71 &  94.12 &  71.76 &  59.59 &  90.14 \\
    GBDTModel\_pro process tree min alerts: 1 &  80.71 &   59.66 &  94.96 &  64.88 &  40.34 &  91.33 \\
    GBDTModel\_pro process tree min alerts: 2 &  80.71 &   59.66 &  94.96 &  64.87 &  40.30 &  91.33 \\
    GBDTModel\_pro process tree min alerts: 3 &  80.71 &   59.66 &  94.96 &  64.87 &  40.30 &  91.33 \\
    GBDTModel\_pro process tree min alerts: 4 &  80.71 &   59.66 &  94.96 &  64.87 &  40.30 &  91.33 \\
  GBDTModel\_pro rolling mean window: 2 &  84.68 &   79.83 &  88.24 &  75.08 &  71.60 &  85.91 \\
  GBDTModel\_pro rolling mean window: 3 &  84.08 &   80.67 &  86.55 &  74.99 &  71.71 &  85.64 \\
  GBDTModel\_pro rolling mean window: 4 &  84.39 &   84.87 &  84.03 &  73.91 &  76.05 &  79.84 \\
  GBDTModel\_pro sum alerts min: 2 &  85.48 &   84.03 &  86.55 &  74.50 &  74.40 &  82.33 \\
  GBDTModel\_pro sum alerts min: 3 &  85.11 &   86.55 &  84.03 &  72.35 &  76.77 &  76.59 \\
  GBDTModel\_pro sum alerts min: 4 &  83.84 &   88.24 &  80.67 &  70.17 &  78.38 &  71.71 \\
GBDTModel\_pro\_tree &  79.02 &   59.09 &  94.96 &  71.08 &  46.68 &  90.51 \\
  GBDTRegression\_pro\_process &  89.71 &   87.39 &  91.60 &  71.84 &  80.57 &  72.52 \\
  MLPModel\_glo\_pro &  66.67 &   13.45 &  93.28 &  57.92 &  19.00 &  90.68 \\
MLPModel\_glo\_pro mean process tree &  67.48 &   16.81 &  93.28 &  59.79 &  25.74 &  90.51 \\
MLPModel\_glo\_pro process tree min alerts: 1 &  67.46 &   13.45 &  94.96 &  57.61 &  17.64 &  90.84 \\
MLPModel\_glo\_pro process tree min alerts: 2 &  67.46 &   13.45 &  94.96 &  57.61 &  17.64 &  90.84 \\
MLPModel\_glo\_pro process tree min alerts: 3 &  67.46 &   13.45 &  94.96 &  57.61 &  17.64 &  90.84 \\
MLPModel\_glo\_pro process tree min alerts: 4 &  67.46 &   13.45 &  94.96 &  57.61 &  17.64 &  90.84 \\
    MLPModel\_glo\_pro rolling mean window: 2 &  67.96 &   28.57 &  88.24 &  58.73 &  32.20 &  84.17 \\
    MLPModel\_glo\_pro rolling mean window: 3 &  67.79 &   34.45 &  84.87 &  58.90 &  34.31 &  83.20 \\
    MLPModel\_glo\_pro rolling mean window: 4 &  68.75 &   41.18 &  83.19 &  58.32 &  40.55 &  78.16 \\
MLPModel\_glo\_pro sum alerts min: 2 &  68.94 &   38.66 &  84.87 &  59.51 &  39.19 &  81.30 \\
MLPModel\_glo\_pro sum alerts min: 3 &  69.04 &   45.38 &  81.51 &  58.47 &  43.17 &  76.80 \\
MLPModel\_glo\_pro sum alerts min: 4 &  70.63 &   53.78 &  79.83 &  57.23 &  47.72 &  71.76 \\
    \end{tabular}
    \caption{Summary of process killing models, validation and test set score metrics [Table 1 of 3]}
    \label{app_tab:paper4_process_killing}
\end{table}

\begin{table}[ht]
\tiny
    \centering
    \begin{tabular}{l|r|r|r|r|r|r}
  & \multicolumn{3}{c}{val} & \multicolumn{3}{c}{test} \\
 model &     f1 &     tnr &    tpr &     f1 &    tnr &    tpr \\\hline
MLPModel\_pro &  71.43 &   56.30 &  79.83 &  57.54 &  52.53 &  69.38 \\
    MLPModel\_pro mean process tree &  72.18 &   57.14 &  80.67 &  57.06 &  53.53 &  67.97 \\
    MLPModel\_pro process tree min alerts: 1 &  72.32 &   54.62 &  82.35 &  57.41 &  49.41 &  71.06 \\
    MLPModel\_pro process tree min alerts: 2 &  72.32 &   54.62 &  82.35 &  57.41 &  49.41 &  71.06 \\
    MLPModel\_pro process tree min alerts: 3 &  72.32 &   54.62 &  82.35 &  57.41 &  49.41 &  71.06 \\
    MLPModel\_pro process tree min alerts: 4 &  72.32 &   54.62 &  82.35 &  57.41 &  49.41 &  71.06 \\
  MLPModel\_pro rolling mean window: 2 &  71.77 &   66.39 &  74.79 &  48.88 &  63.18 &  50.35 \\
  MLPModel\_pro rolling mean window: 3 &  72.65 &   68.91 &  74.79 &  49.23 &  63.71 &  50.57 \\
  MLPModel\_pro rolling mean window: 4 &  72.34 &   73.95 &  71.43 &  46.40 &  67.05 &  45.26 \\
    MLPModel\_pro sum alerts min: 2 &  73.86 &   72.27 &  74.79 &  47.14 &  66.40 &  46.50 \\
    MLPModel\_pro sum alerts min: 3 &  73.68 &   78.99 &  70.59 &  45.27 &  68.12 &  43.36 \\
    MLPModel\_pro sum alerts min: 4 &  73.30 &   82.35 &  68.07 &  44.48 &  69.63 &  41.73 \\
 MLPModel\_pro\_tree &  71.38 &   31.82 &  97.48 &  64.36 &  21.07 &  92.52 \\
   MLPRegression\_pro\_process &  38.89 &   53.78 &  35.29 &  54.83 &  48.73 &  67.05 \\
MLPRegression\_pro\_process mean process tree &  37.32 &   57.14 &  32.77 &  56.75 &  57.37 &  65.15 \\
  NBModel\_glo\_pro &  67.25 &    9.24 &  96.64 &  55.07 &  10.36 &  89.49 \\
 NBModel\_glo\_pro mean process tree &  67.25 &    9.24 &  96.64 &  55.62 &  12.69 &  89.38 \\
  NBModel\_glo\_pro process tree min alerts: 1 &  67.25 &    9.24 &  96.64 &  55.00 &  10.18 &  89.43 \\
  NBModel\_glo\_pro process tree min alerts: 2 &  67.25 &    9.24 &  96.64 &  55.00 &  10.18 &  89.43 \\
  NBModel\_glo\_pro process tree min alerts: 3 &  67.25 &    9.24 &  96.64 &  55.00 &  10.18 &  89.43 \\
  NBModel\_glo\_pro process tree min alerts: 4 &  67.25 &    9.24 &  96.64 &  55.00 &  10.18 &  89.43 \\
NBModel\_glo\_pro rolling mean window: 2 &  67.69 &   19.33 &  92.44 &  55.20 &  15.10 &  87.05 \\
NBModel\_glo\_pro rolling mean window: 3 &  67.73 &   26.05 &  89.08 &  55.32 &  17.32 &  86.02 \\
NBModel\_glo\_pro rolling mean window: 4 &  67.76 &   31.09 &  86.55 &  54.28 &  21.08 &  81.68 \\
 NBModel\_glo\_pro sum alerts min: 2 &  68.17 &   27.73 &  89.08 &  55.70 &  19.40 &  85.64 \\
 NBModel\_glo\_pro sum alerts min: 3 &  67.99 &   31.93 &  86.55 &  54.42 &  22.27 &  81.30 \\
 NBModel\_glo\_pro sum alerts min: 4 &  68.03 &   36.97 &  84.03 &  51.32 &  25.39 &  73.44 \\
NBModel\_pro &  67.06 &    8.40 &  96.64 &  55.60 &   7.03 &  92.63 \\
     NBModel\_pro mean process tree &  67.06 &    8.40 &  96.64 &  56.17 &   9.61 &  92.41 \\
NBModel\_pro process tree min alerts: 1 &  67.06 &    8.40 &  96.64 &  55.56 &   6.78 &  92.68 \\
NBModel\_pro process tree min alerts: 2 &  67.06 &    8.40 &  96.64 &  55.56 &   6.78 &  92.68 \\
NBModel\_pro process tree min alerts: 3 &  67.06 &    8.40 &  96.64 &  55.56 &   6.78 &  92.68 \\
NBModel\_pro process tree min alerts: 4 &  67.06 &    8.40 &  96.64 &  55.56 &   6.78 &  92.68 \\
NBModel\_pro rolling mean window: 2 &  67.69 &   19.33 &  92.44 &  56.01 &  13.41 &  89.81 \\
NBModel\_pro rolling mean window: 3 &  67.52 &   25.21 &  89.08 &  56.18 &  15.81 &  88.78 \\
NBModel\_pro rolling mean window: 4 &  67.99 &   31.93 &  86.55 &  54.94 &  19.61 &  83.90 \\
     NBModel\_pro sum alerts min: 2 &  68.61 &   29.41 &  89.08 &  56.52 &  18.14 &  88.13 \\
     NBModel\_pro sum alerts min: 3 &  68.21 &   32.77 &  86.55 &  55.27 &  21.30 &  83.63 \\
     NBModel\_pro sum alerts min: 4 &  67.81 &   37.82 &  83.19 &  52.25 &  24.42 &  75.77 \\
  NBModel\_pro\_tree &  66.10 &   10.61 &  98.32 &  61.25 &   8.63 &  92.69 \\
NBModel\_pro\_tree mean process tree &  66.10 &   10.61 &  98.32 &  61.25 &   8.63 &  92.69 \\
  RFModel\_glo\_pro &  89.68 &   83.19 &  94.96 &  76.43 &  67.19 &  92.52 \\
 RFModel\_glo\_pro mean process tree &  90.48 &   84.03 &  95.80 &  76.34 &  67.66 &  91.92 \\
  RFModel\_glo\_pro process tree min alerts: 1 &  90.91 &   84.03 &  96.64 &  69.45 &  50.56 &  92.95 \\
  RFModel\_glo\_pro process tree min alerts: 2 &  90.91 &   84.03 &  96.64 &  69.45 &  50.56 &  92.95 \\
  RFModel\_glo\_pro process tree min alerts: 3 &  90.91 &   84.03 &  96.64 &  69.45 &  50.56 &  92.95 \\
  RFModel\_glo\_pro process tree min alerts: 4 &  90.91 &   84.03 &  96.64 &  69.45 &  50.56 &  92.95 \\
RFModel\_glo\_pro rolling mean window: 2 &  92.70 &   94.96 &  90.76 &  80.77 &  78.88 &  89.38 \\
RFModel\_glo\_pro rolling mean window: 3 &  91.30 &   94.96 &  88.24 &  80.19 &  78.67 &  88.51 \\
RFModel\_glo\_pro rolling mean window: 4 &  90.27 &   95.80 &  85.71 &  79.86 &  82.86 &  83.69 \\
 RFModel\_glo\_pro sum alerts min: 2 &  91.30 &   94.96 &  88.24 &  81.50 &  81.53 &  87.97 \\
 RFModel\_glo\_pro sum alerts min: 3 &  90.27 &   95.80 &  85.71 &  79.99 &  84.01 &  82.76 \\
 RFModel\_glo\_pro sum alerts min: 4 &  88.79 &   95.80 &  83.19 &  76.11 &  85.37 &  75.01 \\
RFModel\_pro &  92.37 &   87.39 &  96.64 &  74.57 &  62.71 &  92.95 \\
     RFModel\_pro mean process tree &  92.74 &   88.24 &  96.64 &  74.79 &  64.04 &  92.20 \\
RFModel\_pro process tree min alerts: 1 &  92.74 &   88.24 &  96.64 &  68.75 &  48.23 &  93.39 \\
RFModel\_pro process tree min alerts: 2 &  92.74 &   88.24 &  96.64 &  68.75 &  48.23 &  93.39 \\
RFModel\_pro process tree min alerts: 3 &  92.74 &   88.24 &  96.64 &  68.75 &  48.23 &  93.39 \\
RFModel\_pro process tree min alerts: 4 &  92.74 &   88.24 &  96.64 &  68.75 &  48.23 &  93.39 \\
RFModel\_pro rolling mean window: 2 &  93.22 &   94.12 &  92.44 &  78.28 &  73.83 &  89.76 \\
RFModel\_pro rolling mean window: 3 &  91.38 &   94.12 &  89.08 &  77.47 &  73.25 &  88.78 \\
RFModel\_pro rolling mean window: 4 &  89.96 &   94.12 &  86.55 &  77.08 &  77.70 &  83.85 \\
     RFModel\_pro sum alerts min: 2 &  91.38 &   94.12 &  89.08 &  77.98 &  74.65 &  88.40 \\
     RFModel\_pro sum alerts min: 3 &  89.96 &   94.12 &  86.55 &  77.05 &  77.52 &  83.96 \\
     RFModel\_pro sum alerts min: 4 &  88.50 &   94.12 &  84.03 &  73.11 &  79.35 &  75.61 \\
  RFModel\_pro\_tree &  90.35 &   82.58 &  98.32 &  74.20 &  52.44 &  92.74 \\
    RFRegression\_pro\_process &  91.94 &   87.39 &  95.80 &  74.77 &  66.05 &  90.35 \\
  SVMModel\_glo\_pro &  65.23 &   15.97 &  89.08 &  57.34 &  24.24 &  86.23 \\
SVMModel\_glo\_pro mean process tree &  65.23 &   15.97 &  89.08 &  58.11 &  27.39 &  85.91 \\
SVMModel\_glo\_pro process tree min alerts: 1 &  65.23 &   15.97 &  89.08 &  57.32 &  23.81 &  86.45 \\
SVMModel\_glo\_pro process tree min alerts: 2 &  65.23 &   15.97 &  89.08 &  57.32 &  23.81 &  86.45 \\
SVMModel\_glo\_pro process tree min alerts: 3 &  65.23 &   15.97 &  89.08 &  57.32 &  23.81 &  86.45 \\
SVMModel\_glo\_pro process tree min alerts: 4 &  65.23 &   15.97 &  89.08 &  57.32 &  23.81 &  86.45 \\
    SVMModel\_glo\_pro rolling mean window: 2 &  65.15 &   26.05 &  84.03 &  57.98 &  33.52 &  81.84 \\
    SVMModel\_glo\_pro rolling mean window: 3 &  64.65 &   31.09 &  80.67 &  58.14 &  35.46 &  80.98 \\
    SVMModel\_glo\_pro rolling mean window: 4 &  64.31 &   38.66 &  76.47 &  56.76 &  40.37 &  75.34 \\
SVMModel\_glo\_pro sum alerts min: 2 &  65.05 &   36.13 &  78.99 &  58.35 &  39.08 &  79.13 \\
SVMModel\_glo\_pro sum alerts min: 3 &  64.75 &   42.02 &  75.63 &  57.05 &  43.24 &  74.15 \\
SVMModel\_glo\_pro sum alerts min: 4 &  64.89 &   51.26 &  71.43 &  54.70 &  47.40 &  67.59 \\
SVMModel\_pro &  66.47 &    5.88 &  96.64 &  56.92 &  10.33 &  93.71 \\
    \end{tabular}
    \caption{Summary of process killing models, validation and test set score metrics [Table 2 of 3]}
    \label{app_tab:chapter4_process_killing_2}
\end{table}

\begin{table}[ht]
\tiny
    \centering
    \begin{tabular}{l|r|r|r|r|r|r}
  & \multicolumn{3}{c}{val} & \multicolumn{3}{c}{test} \\
    SVMModel\_pro mean process tree &  67.25 &    9.24 &  96.64 &  57.55 &  13.34 &  93.33 \\
    SVMModel\_pro process tree min alerts: 1 &  66.47 &    5.88 &  96.64 &  56.21 &   7.28 &  93.88 \\
    SVMModel\_pro process tree min alerts: 2 &  66.47 &    5.88 &  96.64 &  56.21 &   7.28 &  93.88 \\
    SVMModel\_pro process tree min alerts: 3 &  66.47 &    5.88 &  96.64 &  56.21 &   7.28 &  93.88 \\
    SVMModel\_pro process tree min alerts: 4 &  66.47 &    5.88 &  96.64 &  56.21 &   7.28 &  93.88 \\
  SVMModel\_pro rolling mean window: 2 &  66.87 &   15.97 &  92.44 &  58.60 &  22.02 &  90.30 \\
  SVMModel\_pro rolling mean window: 3 &  67.30 &   24.37 &  89.08 &  58.82 &  24.42 &  89.27 \\
  SVMModel\_pro rolling mean window: 4 &  67.99 &   31.93 &  86.55 &  57.98 &  28.97 &  84.66 \\
    SVMModel\_pro sum alerts min: 2 &  67.96 &   28.57 &  88.24 &  59.52 &  27.61 &  88.73 \\
    SVMModel\_pro sum alerts min: 3 &  68.90 &   35.29 &  86.55 &  59.06 &  33.35 &  84.12 \\
    SVMModel\_pro sum alerts min: 4 &  68.75 &   41.18 &  83.19 &  56.68 &  38.87 &  76.10 \\
 SVMModel\_pro\_tree &  65.73 &    9.09 &  98.32 &  61.79 &   9.88 &  93.19 \\
  dqn &  51.71 &   72.27 &  44.54 &  27.74 &  55.50 &  26.94 \\
   random\_search\_glo\_pro\_RNN &  87.69 &   77.31 &  95.80 &  71.83 &  59.63 &  90.24 \\
random\_search\_glo\_pro\_RNN mean process tree &  88.03 &   78.15 &  95.80 &  72.50 &  61.67 &  89.81 \\
  random\_search\_glo\_pro\_RNN\_Regression &  85.71 &   72.27 &  95.80 &  72.44 &  61.78 &  89.59 \\
 random\_search\_pro\_RNN &  91.20 &   85.71 &  95.80 &  72.63 &  59.63 &  91.82 \\
    random\_search\_pro\_RNN mean process tree &  91.20 &   85.71 &  95.80 &  73.03 &  60.92 &  91.49 \\
  random\_search\_pro\_RNN\_Regression &  88.37 &   78.99 &  95.80 &  72.71 &  60.70 &  91.06 \\
  random\_search\_pro\_RNN\_tree &  88.19 &   80.67 &  94.12 &  73.72 &  65.79 &  88.56 \\
    \end{tabular}
    \caption{Summary of process killing models, validation and test set score metrics [Table 3 of 3]}
    \label{app_tab:chapter4_process_killing_3}
\end{table}

\end{document}